\documentclass{aa}  
\usepackage{txfonts}
\usepackage{graphicx}
\usepackage{natbib}
\usepackage{multirow} 
\usepackage{lscape}
\usepackage{longtable}
\usepackage{amssymb,amsmath}
\usepackage{color}
\usepackage{hyperref}
\bibpunct{(}{)}{;}{a}{}{,} 
\newcommand{\micron}{\mbox{$\mu$m}}

\defcitealias{Dionatos:17a}{Paper I}

\begin{document}

   \title{Feedback of molecular outflows from protostars in NGC 1333}
\subtitle{revealed by Herschel and Spitzer spectro-imaging observations}

   \author{Odysseas Dionatos
          \inst{1}
          \and
    Lars. E. Kristensen\inst{2}
    	\and
	Mario Tafalla\inst{3}       
          \and
          Manuel G\"udel\inst{1}
          \and
          Magnus Persson\inst{4}
                   }

   \institute{Department of Astrophysics, University of Vienna,
              T\"urkenschanzstrasse 17, A-1180 Vienna\\
              \email{odysseas.dionatos@univie.ac.at}
         \and
		Centre for Star and Planet Formation, Niels Bohr Institute and Natural History Museum of Denmark, University of Copenhagen, {\O}ster Voldgade 5-7, 1350 Copenhagen K, Denmark
	\and
		Observatorio Astron{\'o}mico Nacional (IGN), Alfonso XII 3, 28014 Madrid, Spain
	\and
		Department of Space, Earth and Environment, Chalmers University of Technology, Onsala Space Observatory, 439 92, Onsala, Sweden
             }


  \abstract
   {Far infrared cooling of excited gas around protostars has been predominantly studied in the context of pointed observations. Large scale spectral maps of star forming regions enable the simultaneous, comparative study of the gas excitation around an ensemble of sources at a common frame of reference, therefore providing direct insights in the multitude of physical processes involved. }
   {We employ extended spectral-line maps to decipher the excitation, the kinematical and dynamical processes in NGC\,1333 as revealed by a number of different molecular and atomic lines, aiming to set a reference for the applicability and limitations of difference tracers in constraining particular physical processes.}
   {We reconstruct line maps for H$_2$, CO, H$_2$O and [\ion{C}{I}] using data obtained with the Spitzer/IRS spectrograph and the Herschel HIFI and SPIRE instruments. We compare  the morphological features revealed in the maps and derive the gas excitation conditions for regions of interest employing LTE and non-LTE methods. We also calculate the kinematical and dynamical properties for each outflow tracer in a consistent manner, for all observed outflows driven by protostars in NGC\,1333. We finally measure the water abundance in outflows with respect to carbon monoxide and molecular hydrogen.}
   {CO and H$_2$ are highly excited around B-stars and at lower levels trace protostellar outflows. H$_2$O emission is dominated by a moderately fast component associated with outflows.  H$_2$O also displays a weak, narrow-line component in the vicinity of B-stars associated to their UV field. This narrow component it is also present in a handful of outflows, indicating UV radiation generated in shocks. Intermediate J CO lines appear brightest at the locations traced by the narrow H$_2$O component, indicating that beyond the dominating collisional processes, a secondary, radiative excitation component can also be active. The morphology, kinematics, excitation and abundance variations of water are consistent with its excitation and partial dissociation in shocks. Water abundance ranges between 5 $\times$ 10$^{-7}$ and $\sim$ 10$^{-5}$, with the lower values being more representative. Water is brightest and most abundant around IRAS\,4A which is consistent with the latter hosting a hot corino source. [\ion{C}{I}] traces dense and warm gas in the envelopes surrounding protostars. Outflow mass flux is highest for CO and decreases by one and two orders of magnitude for H$_2$ and H$_2$O, respectively.} 
   {Large scale spectral line maps can provide unique insights into the excitation of gas in star-forming regions. Comparative analysis of line excitation and morphologies at different locations allows to decipher the dominant excitation conditions in each region but also isolate exceptional cases. } 


   \maketitle
%

\begin{figure*}[!ht]
\centering
\includegraphics[width=17.2cm]{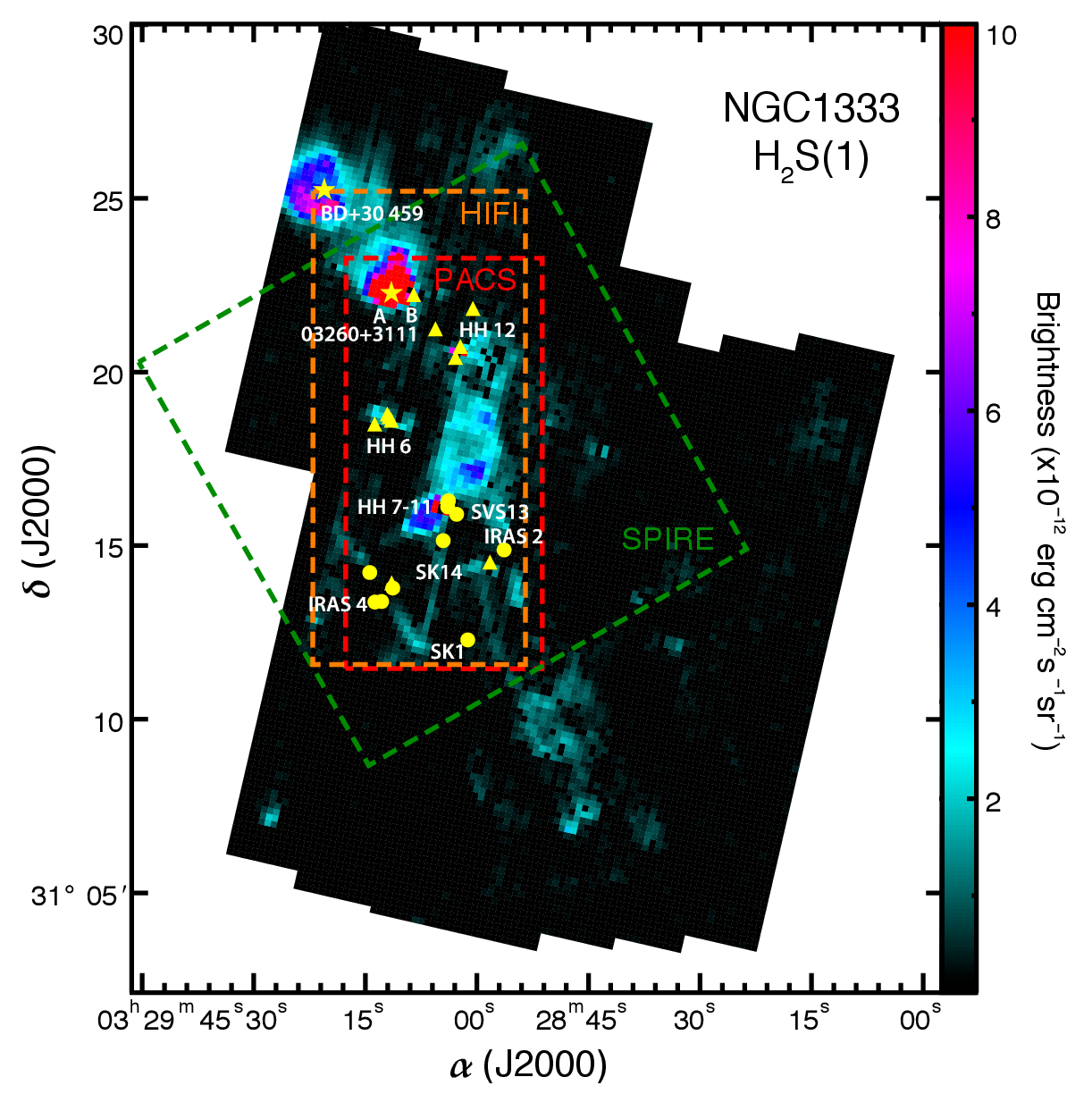}
\caption {Spectral line map of the H$_2$ S(1) pure rotational transition observed with Spitzer/IRS tracing gas excited in outflows and photon-dominated regions in NGC\,1333. Positions of Class 0 and I sources in the star-forming region are indicated with filled circles and triangles, respectively. Filled star-symbols mark the position of  the B stars 03269+3111 and BD+30459 in the field. Dashed frames delineate the extent of the regions mapped with Herschel HIFI and SPIRE that are discussed in the following sections. The extent of the PACS maps presented in \citetalias{Dionatos:17a} is also shown for reference.} 
\label{fig:1}
\end{figure*}

\section{Introduction}

Since the discovery of Herbig-Haro objects, emission lines tracing protostellar jets and outflows have acted as  beacons witnessing the formation of stars. Over the course of the last couple of decades a series of infrared and millimeter/submillimeter facilities with increasing sensitivity  enabled the detection of many new lines from increasingly complex molecular species. These detections testify that the surroundings of protostars are not only dynamically active, but also chemically complex. The study of emission lines provides us with means to understand the physical, kinematical, and dynamical conditions of the emitting gas. When combined with spatial information in spectral line maps they become essential tools to decipher the complex interplay between different physical components that lead to the formation of stars.

Emission lines provide efficient cooling of gas that can be excited through various energetic processes such as the interaction with energetic radiation or  the action of supersonic shocks. Lines emitted  in the mid- and far-infrared part of the spectrum provide efficient cooling mechanisms of the gas in dense environments around protostars where extinction from the protostellar envelope is very high. Surveys of the environment around protostars with infrared space facilities such as the Herschel Space Observatory and the Spitzer Space telescope, have revealed that the major coolants of excited gas are lines from molecules  (H$_2$ CO, H$_2$O), atoms (O, S) and ions (Fe$^+$, Si$^+$) \citep[e.g.,][]{Karska:18a, Green:13a}. The contribution of each line in different processes however remains unclear and it is hindered by instrumental limitations in terms of sensitivity, angular and spectral resolution, and spatial coverage. 

 Aiming to observe a large sample of sources during the limited lifetime of cryogenically cooled space missions, spectra are taken with a single observation, pointed on the source. In a few cases, small scale maps covering the protostar along with its outflows have provided valuable information on the great diversity of excitation mechanisms that can be revealed by the spatial association of different molecular and atomic tracers. Such an example is the case of HH\,211 that was mapped with Spitzer/IRS and Herschel/PACS spectrographs (\citet{Tappe:09a, Dionatos:10a} and \citet{Dionatos:18a}, respectively), where the morphological diversity of the emission patterns from different molecules have helped us to backtrace the prevailing excitation processes.

Large scale spectral maps of star forming regions are even more scarce However such observations can help us study star-formation consistently for a number of individual objects and at the same time the influence of the assembly of forming stars onto their environment. The only case for which such maps exist in the mid- and far-IR  are for the NGC\,1333 star forming region that has been observed with Spitzer/IRS \citep{Maret:09a} and Herschel/PACS \citep{Dionatos:17a}. In this paper we present Herschel HIFI and SPIRE maps of NGC\,1333 complemented with additional Spitzer/IRS maps not included in the original work by  \citet{Maret:09a}, aiming to understand kinematic and energetic properties of protostellar jets and outflows. The paper is structured as follows. In Sect.~\ref{sec:2} we report on the reduction of the data and in Sect.~\ref{sec:3} we discuss the morphology of the reconstructed emission line maps. Analysis of the data is then discussed  Sect.~\ref{sec:4}, where we derive the excitation conditions and the outflow kinematical and dynamical properties, along with estimations of the water abundance. In Sect.~\ref{sec:5} we discuss our results, aiming to set a reference for the usage and limitations of of different tracers in constraining the jet and outflow properties. Finally, a summary is provided in Sect.~\ref{sec:6}







   

\section{Observations and data reduction} \label{sec:2}

\subsection{Spitzer IRS}

 Spitzer maps employed in this work belong to two different observing programs with identification numbers (ID) 116 and 20378. Observations were obtained on the 6th of March 2006 and between March and September 2006 for ID 116 and 20378, respectively. Data were retrieved from the Spitzer Heritage Archive (SHA\footnote[1]{https://sha.ipac.caltech.edu}). Observations from program 116 employ the long-wavelength, low-resolution (LL, 13.9-39.9\,$\micron$) module of the Infrared Spectrograph (IRS) in slit-scan mode to cover an area of 23.96$\arcmin\times$ 17.61$\arcmin$ and are presented in this work for the first time (see Fig.~\ref{fig:1}). Observations from program 20378 employ the short-wavelength, low-resolution (SL, 5.1-14.2\,$\micron$), short-wavelength high resolution (SH, 9.9-19.5\,$\micron$) and long-wavelength high-resolution (LH, 18.8-37.1\,$\micron$) IRS modules have been presented and discussed in \citet{Maret:09a}. SL module maps cover an area of 13.9 $\arcmin\times$ 9.1$\arcmin$, while the SH and LH cover 9.86 $\arcmin\times$ 5.91$\arcmin$ and 9.36 $\arcmin\times$ 8.47$\arcmin$, respectively.  The high resolution IRS modules can produce higher quality maps compared to the low resolution ones. However the SH maps leave an unobserved gap at the center of the cluster and  provide a rather limited coverage. The LH maps on the other hand, with the exception of the central cluster region are sparse and do not provide a complete spatial coverage \citep[see][]{Maret:09a}. For those reasons we based our analysis mostly on the low-resolution modules and used the high resolution ones in regions where they could provide complete coverage. All data were initially reduced using the CUBISM software \citep{Smith:07a} and rogue pixels were removed using spectra from emission-free areas of the maps. Further processing to produce line maps included gaussian-line fitting after removing a first or second order polynomial baseline.

  \begin{figure}[!ht]
\centering
\includegraphics[width=7.5cm]{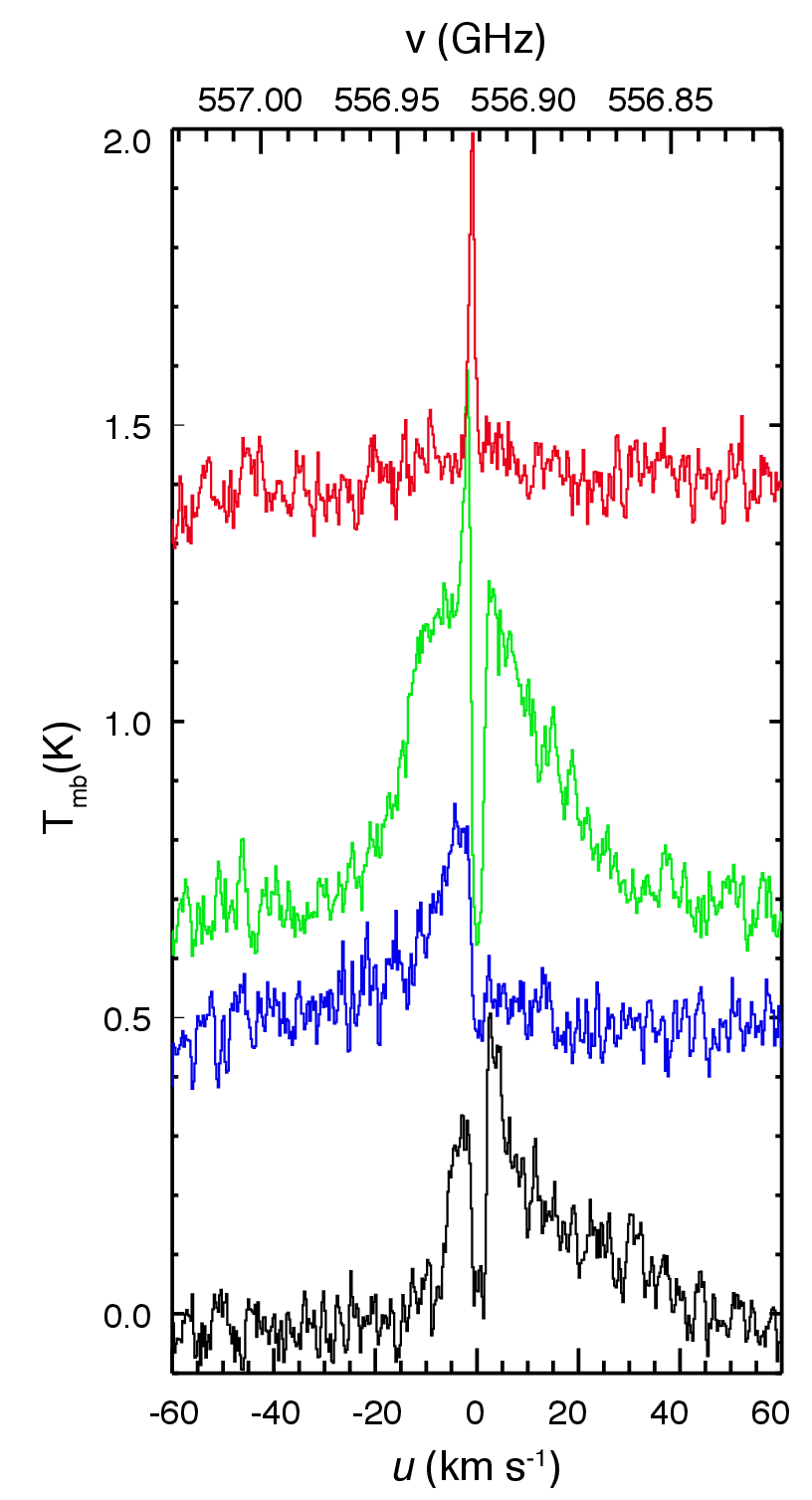}
\caption {Characteristic examples of H$_2$O $1_{10} - 1_{01}$ line profiles from different regions across NGC\,1333 observed with HIFI. Velocity scale is shifted so that the source velocity is at 0~km~s$^{-1}$ From bottom to top: red-shifted lobe to the north of IRAS\,4A/B complex (black), blue-shifted emission to the south of IRAS\,2A (blue), on-source emission from IRAS\,4A (green), and from the surroundings of the B-star 03260+3111 to the north (red).} 
\label{fig:2}
\end{figure}

\subsection{Herschel HIFI }

 Observations were retrieved from the Herschel Science Archive (HSA\footnote[2]{http://archives.esac.esa.int/hsa/whsa/}) and were obtained on the 2nd of February 2012. The Heterodyne Instrument for the Far-Infrared \citep[HIFI;][]{deGraauw:10a} onboard the Herschel Space Observatory \citep{Pilbratt:10a}  was used in on-the-fly (OTF) mapping mode to scan the star-forming region NGC\,1333 at Nyquist spatial sampling. From the Wide-Band Spectrometer (WBS) back-end providing a $\sim$4 GHz bandwidth the sub-band 1a was used  to observe the H$_2$O line at 557 GHz. The resulting map covers a region of 13.64$\arcmin\times$ 6.48$\arcmin$ while the pixel size at the frequency of the H$_2$O line at 557 GHz is 18.6$\arcsec$. The corresponding instrumental half-power beam width is $\sim$37.5$\arcsec$. HIFI data were reduced using HIPE version 15.0.3244 using calibration file HIFI\_CAL\_25\_0. HIFI line maps were created using home-grown pipelines. In these, the line intensity and line-centroid velocity shifts for each spaxel were determined through fitting a gaussian after subtracting a first- or second-order polynomial baseline. The coverage of the HIFI maps is presented in Fig.~\ref{fig:1}.


 \begin{figure*}[!ht]
\centering
\includegraphics[width=\hsize]{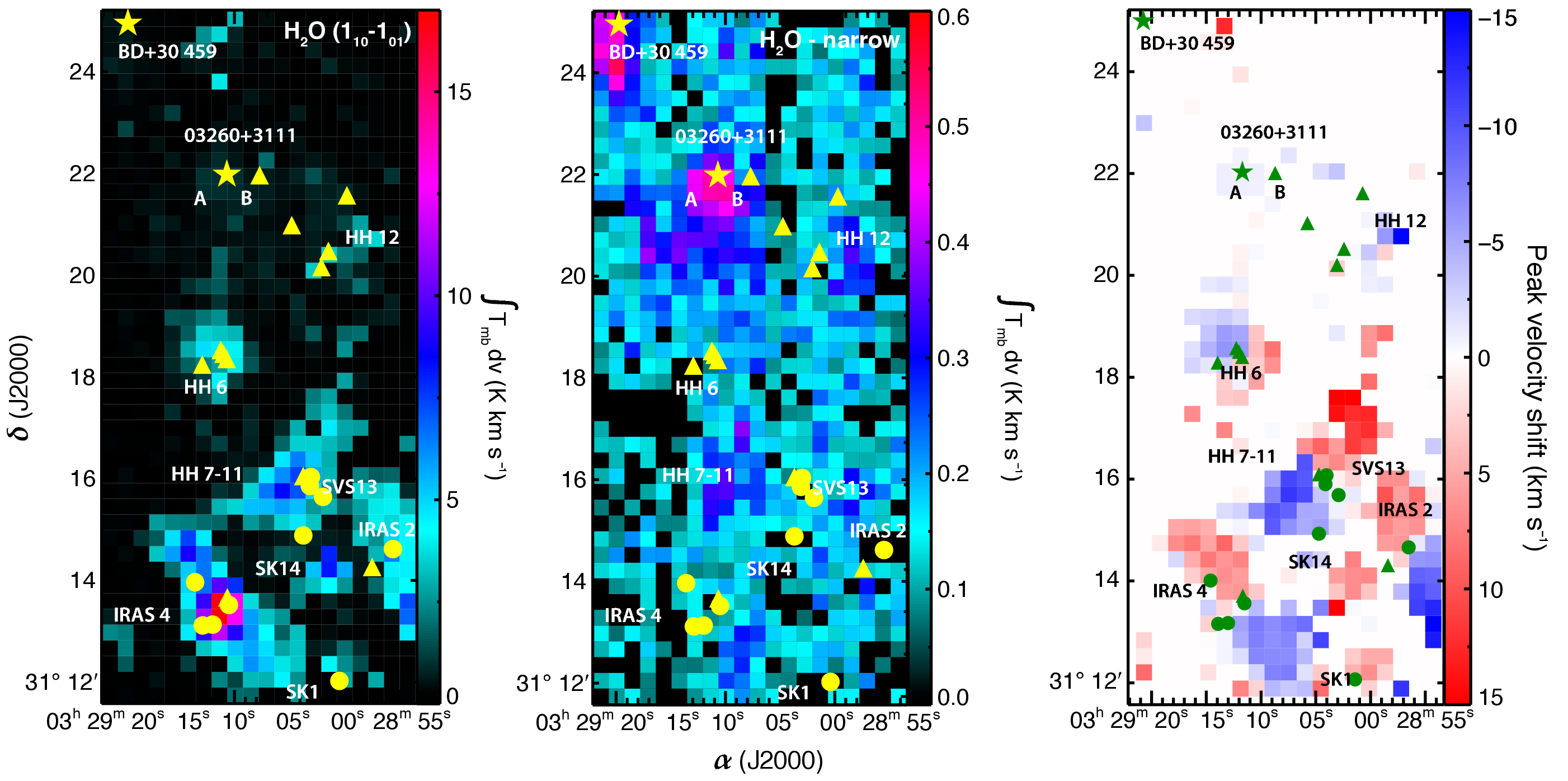}
\caption {HIFI maps of the broad and narrow $1_{10} - 1_{01}$ H$_2$O line components (left and central panels, respectively). The right-hand panel presents a velocity map corresponding to the gaussian centroid-shift for the broad component that reflects the motion for the bulk of the outflowing gas.} 
\label{fig:3}
\end{figure*}

\subsection{Herschel SPIRE }

 Observations were obtained between  August 27th and 28th  2013 and data records were also retrieved from the HSA. The Spectral and Photometric Imaging Receiver  \citep[SPIRE;][]{Griffin:10a} onboard the Herschel Space Observatory \citep{Pilbratt:10a}  was used in spectral raster mode at  the intermediate sampling rate resulting in one beam spatial sampling. The SPIRE spectrograph consists of two modules, the Spectrograph Short Wavelengths (SSW) module covering from 191 to 318\,$\micron$ and the Spectrograph Long Wavelengths (SLW) module ranging between 294 and 671\,$\micron$.  The sampling rate results in spectral maps with a spaxel size of 19$\arcsec$ and 35$\arcsec$ for the SSW and SLW modules, respectively, while the total coverage of the combined raster maps extends to an area of  $\sim$ 13$\arcmin\times$ 13$\arcmin$, shown in Fig.~\ref{fig:1}. SPIRE data were reduced in HIPE version 15.0.3244 using the standard pipeline for extended sources with the SPIRE calibration dataset spire\_cal\_13\_1, including apodization. Beyond basic reduction, SPIRE data were further processed with home-grown routines. As with the other data, line maps were constructed by fitting a gaussian to each line of interest after subtracting a first order polynomial baseline.

 \section{Line-emission morphology}
 \label{sec:3}

\subsection{IRS maps}

In Fig.\ref{fig:1} we present the angular distribution of the H$_2$ S(1) rotational transition at 17\,$\micron$. The H$_2$ map has a much larger coverage and provides higher contrast when compared to the maps obtained with the high resolution IRS modules of the same line \citep[see][]{Maret:09a}. In addition, spectral-line maps are not affected by continuum and scattered light emission as seen in the IRAC bands \citep{Gutermuth:08a} providing a clearer view of the excited gas. Morphologically, the H$_2$ emission peaks at the position of the B star 03260+3111 to the north-east and has a secondary peak surrounding BD\,+30459 that resembles  an extended shell. At lower levels, molecular hydrogen delineates most known protostellar outflows, with the ones  clustered around the SVS13 system being the most prominent. The morphology of the H$_2$ emission is very similar to  that of the  [\ion{O}{i}] maps presented in \citet{Dionatos:17a} (hereafter Paper I) indicating that most of the molecular hydrogen is excited in the same shock interfaces as atomic oxygen. Finally, the  H$_2$ emission extends far to the south, beyond the regions covered with any of the Herschel instruments discussed in the following paragraphs.

\subsection{HIFI maps} \label{sec:3.2}

The profile of the H$_2$O $1_{10} - 1_{01}$ line presents a wide morphological diversity when observed at different locations of NGC\,1333. Four representative line profiles are shown in Fig.~\ref{fig:2}. The lower two spectra at the bottom of Fig.~\ref{fig:2} present wide, high-velocity line-wings extending to velocities of $\pm$\,50\,km\,s$^{-1}$. In striking contrast, the emission line-profile at the top of Fig.~\ref{fig:2}  is very narrow with a width of just 2-3 km s$^{-1}$ residing at the velocity of local standard of rest of the cloud ($v_{lsr}$). The second spectrum below the top appears to be combining both high velocity wings and a very narrow peak combined with a very narrow absorption feature close to $v_{lsr}$. The wide morphological diversity of the $1_{10} - 1_{01}$ H$_2$O line  at different locations of NGC\,1333 was first reported based  on observations with the Submillimeter Wave Astronomy Satellite (SWAS) \citep{Bergin:03a}.  Similar morphological characteristics based on Herschel pointed observations around the protostellar sources IRAS\,2A, IRAS\,4A and IRAS\,4B have also been reported in \citet{Kristensen:12a}  and \citet{Mottram:14a}, where these authors further decompose the high velocity wings into a broad and a medium velocity components.

The integrated intensity map of the H$_2$O  $1_{10} - 1_{01}$ line is presented in the left panel of Fig.~\ref{fig:3}. Water emission appears to bright at the center and the south of NGC\,1333, tracing the general outflow morphology as recovered in the H$_2$ maps of Fig.~\ref{fig:1} but also the CO maps of  \citet{Plunkett:13a}, and the jet morphology revealed by [\ion{O}{i}] \citetalias{Dionatos:17a}.  To the north, water does not follow the emission pattern seen in the maps of all other molecules and atoms, which peak around the B stars to the north-east, and instead shows intensity maxima in the surroundings of the protostellar system IRAS\,4A/B to the south-east.  When mapping the emission corresponding only to the very narrow component (Fig.~\ref{fig:2}), the morphology changes drastically. Presented in the central panel of Fig.~\ref{fig:3}, the narrow component peaks in the vicinity of the B-stars to the north and shows secondary maxima 

\begin{landscape}
\begin{figure}
\centering
\resizebox{\hsize}{!}{
\includegraphics{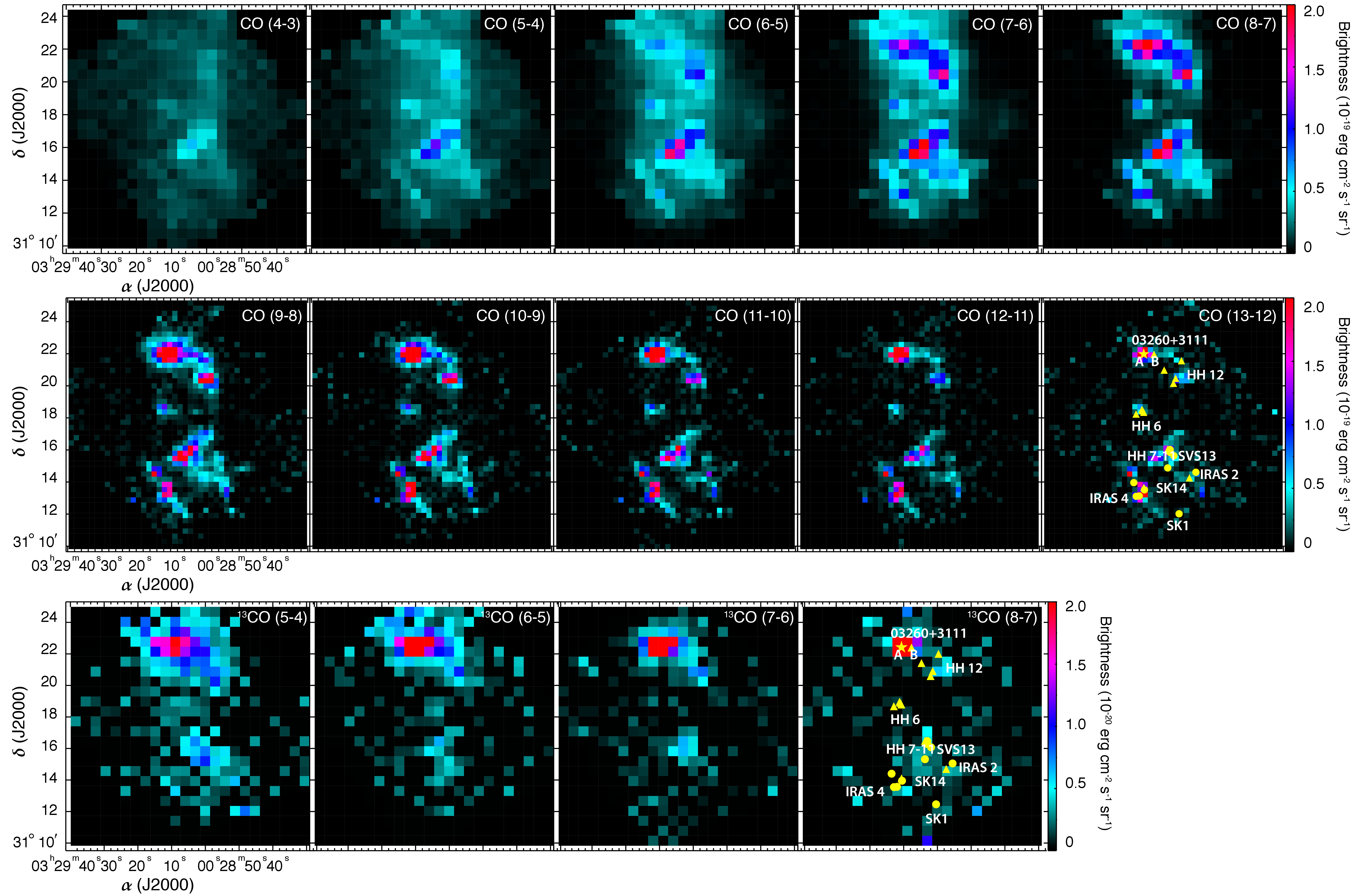}}
\caption{SPIRE spectral-line maps of the $^{12}$CO ladder observed with the long- and short-wavelength modules (upper and central panels, respectively). $^{13}$ CO maps observed with the long-wavelength module are presented at the lower panel.} 
\label{fig:4}
\end{figure}
\end{landscape}

\noindent around the Herbig-Haro objects HH\,7-11 and HH\,12, in line with the emission pattern traced in the line maps of other species. It is worth noting that the narrow component around the IRAS\,4A/B system, appears to be rather weak despite it being present in the line profile (Fig.~\ref{fig:2}). The narrow component represents a small fraction of the total water emission and the peak intensity in the narrow-component maps is more than an order of magnitude lower than the maximum intensity of the fast component. 
While the water lines observed with HIFI are well resolved, their compound profiles make it complex to retrieve information on the overall gas motions. To this end, we fit the water lines with gaussian curves and calculate the velocity shift of the gaussian centroids, as this method provides to a high accuracy the motions for the bulk of the gas in outflows \citepalias[see also][for a detailed description of the method]{Dionatos:17a}.

In the right hand panel of Fig.~\ref{fig:3} we present the velocity-shift map based on the the gaussian-line centroids for the broad line component. Peak velocities for the bulk of the gas reach $\sim\pm$15~km~s$^{-1}$, however maxima do not seem correlate with the peak of the broad H$_2$O emission shown in the left-most panel of Fig.~\ref{fig:3}.  When emission is binned in velocity channel maps (Fig.~\ref{onlinefig:1}), then water in the line-wings can exceed 30~km~s$^{-1}$. At those velocities emission mainly traces fast material in the outflows of IRAS\,4A and IRAS\,2A in the south.

In Fig.~\ref{onlinefig:2} we present the line map of a serendipitous detection of a line centered at 558.9515\,GHz which possibly corresponds to emission from acetone ((CH$_3$)$_2$CO) or methyl-formate (CH$_3$OCHO). The line profile is narrow and integrated emission levels are low and comparable to those reached by \textbf{the} narrow H$_2$O component. Line emisison  peaks at the locations of known protostellar sources (e.g., IRAS4\,A/B, IRAS\,2A, HH\,12) and provides evidence of active  chemistry in the extended envelopes of young stellar objects.


\subsection{SPIRE maps}

In Fig.~\ref{fig:4} we present maps for a series of CO transitions detected with the SPIRE spectrograph. The top row includes $^{12}$CO maps ranging from $J_{up}=4$ to $J_{up}=8$ observed with the SLW module while the middle panel continues with maps for transitions with $J_{up}$ from 9 to 13 detected with the SSW module. The lower panel displays maps of  $^{13}$CO maps with $J_{up}$ between 5 and 8 (at the resolution of the SLW module). Brightness scale is kept constant across all maps of Fig.~\ref{fig:4} in order to facilitate direct comparisons.

\begin{figure}
\includegraphics[width=8.8cm]{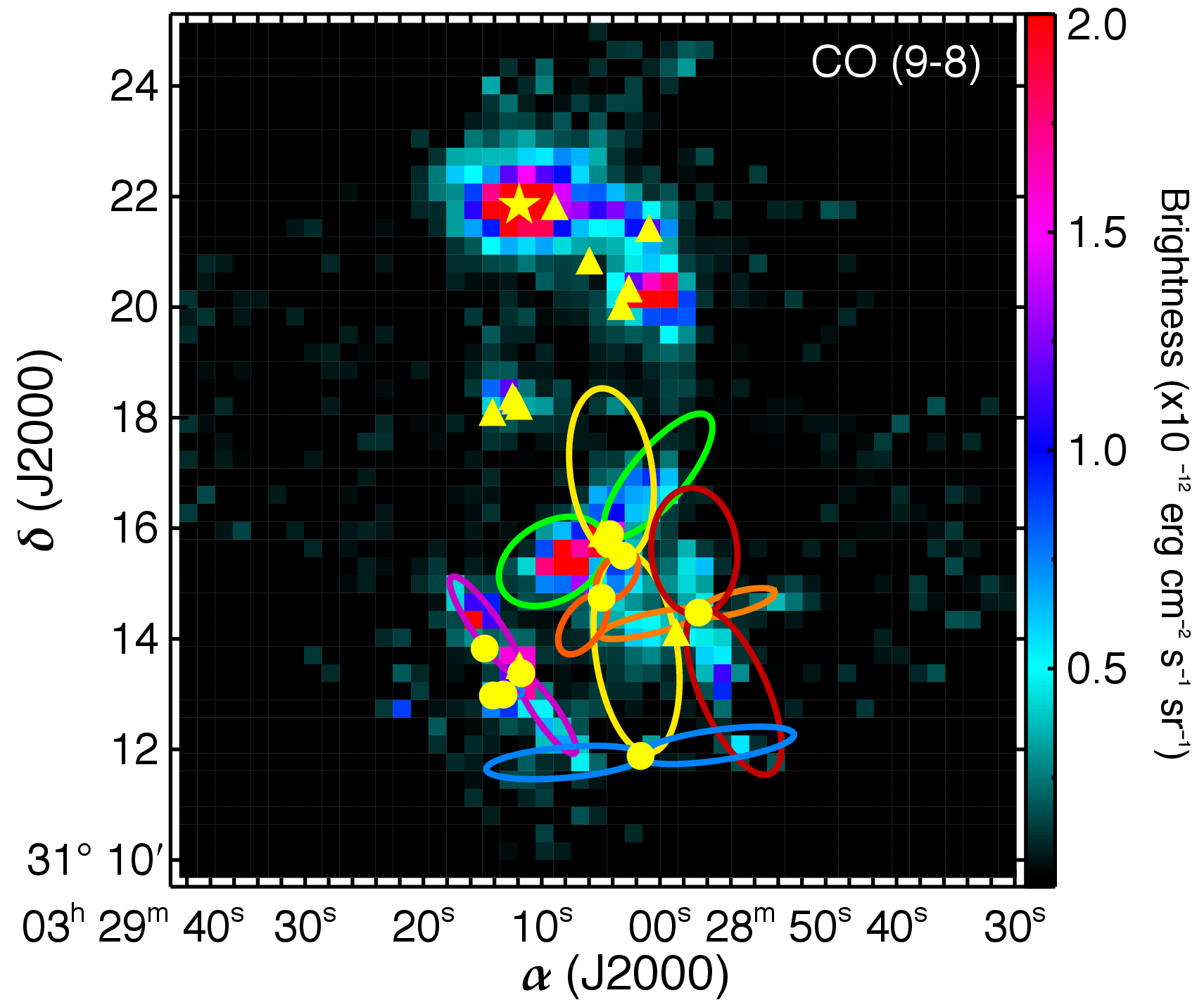}
\caption{SPIRE map with superposition of the elliptical regions adopted from \citet{Plunkett:13a}, which are used to calculate the CO and H$_2$ outflow properties in a consistent manner.} 
\label{fig:6}
\end{figure}

 Lower $J$, $^{12}$CO maps are rather featureless with the exception of the HH\,7-11 outflow which stands out as a bright emission region in all cases. Integrated emission maps at low $J$ CO transitions are dominated by the optically thick component and cannot differentiate the fast moving gas in outflows \citep[e.g.,][]{Dionatos:10b}.  At $J_{up}=7$ two additional peaks of emission become apparent around the B star 03260+3111 in the north and in the vicinity of HH\,12. At $J_{up}=9$ and above more features become increasingly apparent and combined with the higher resolution provided by the SSW module, the shapes of most outflows can readily be recognized (see also Fig.~\ref{fig:6}). 

$^{13}$CO maps possess a very different morphology when compared with the corresponding $^{12}$CO maps of the same upper level. $^{13}$CO shows an outstanding peak of emission around source 03260+3111 with a secondary peak around SVS13. The difference in the morphology revealed by the two CO isotopologues  most likely reflects differences in the  excitation and optical depth that seems to vary significantly between, for example, the B-star 03260+3111 and the Herbig-Haro object HH 7-11. A visual inspection between the maps at the limit where no new structures appear in higher $J$ maps  suggests that most of  the$^{12}$CO emission becomes optically thin at $J_{up}\sim8$. This value is consistent with the findings of \citet{Yang:18a} which are based on detailed radiative transfer models applied on a number of different star-forming regions.  

\begin{figure}
\includegraphics[width=8.8cm]{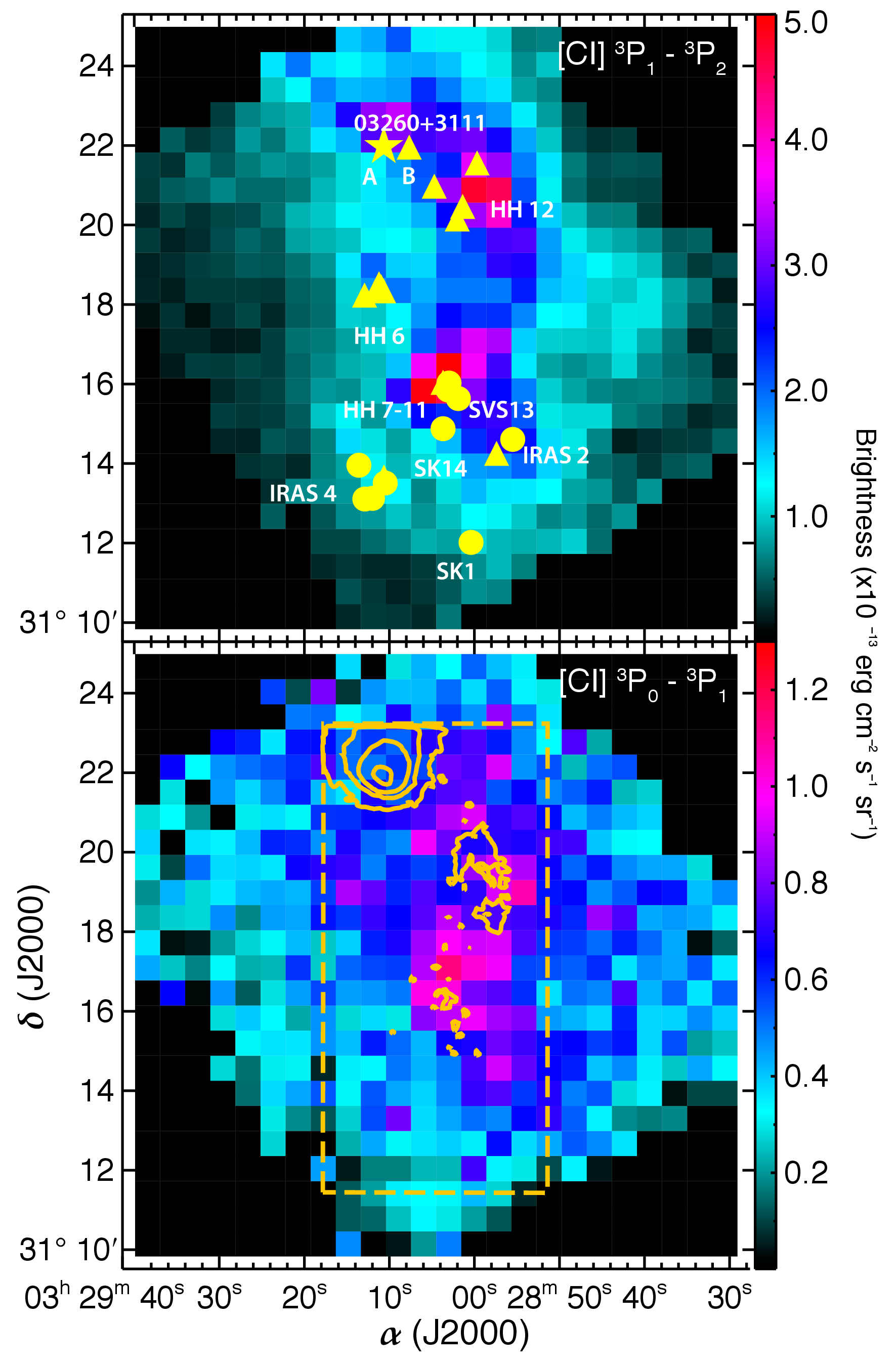}
\caption{SPIRE maps of the lowest energy transitions of [\ion{C}{i}] observed with the long-wavelength module. Upper panel shows the map corresponding to the fundamental transition [\ion{C}{i}] ($^3$P$_0$ - $^3$P$_1$) at 610 $\mu$m while the lower panel displays the map corresponding to the first transition ($^3$P$_1$ - $^3$P$_2$) centered at  370.4 $\mu$m ($\sim$492 and 809\,GHz respectively). To enable comparisons, the [\ion{C}{ii}] emission from \citetalias {Dionatos:17a} is superimposed in the lower panel (orange contours). } 
\label{fig:5}
\end{figure}

In Fig.~\ref{fig:5} we present the distribution of the emission of [\ion{C}{i}]. The lowest energy transition ($^3$P$_0$ - $^3$P$_1$) appears to have a more homogeneous, distributed morphology resembling that of the lowest $J$ CO transitions. The maps of the first excited level ($^3$P$_1$ - $^3$P$_2$) appear to have a more pronounced contrast between the regions around protostars and the surrounding medium. Comparing the  v maps to the distribution of its singly ionized state [\ion{C}{ii}], presented in \citetalias{Dionatos:17a}, we find that the distributions of the two atomic tracers are very different. [\ion{C}{ii}]  is clearly tracing the photon-dominated region around the B-star  03260+3111 whereas [\ion{C}{i}] highlights higher density regions, possibly extended envelopes, surrounding embedded protostars.


\begin{figure*}[!h]
\includegraphics[width=17.9cm]{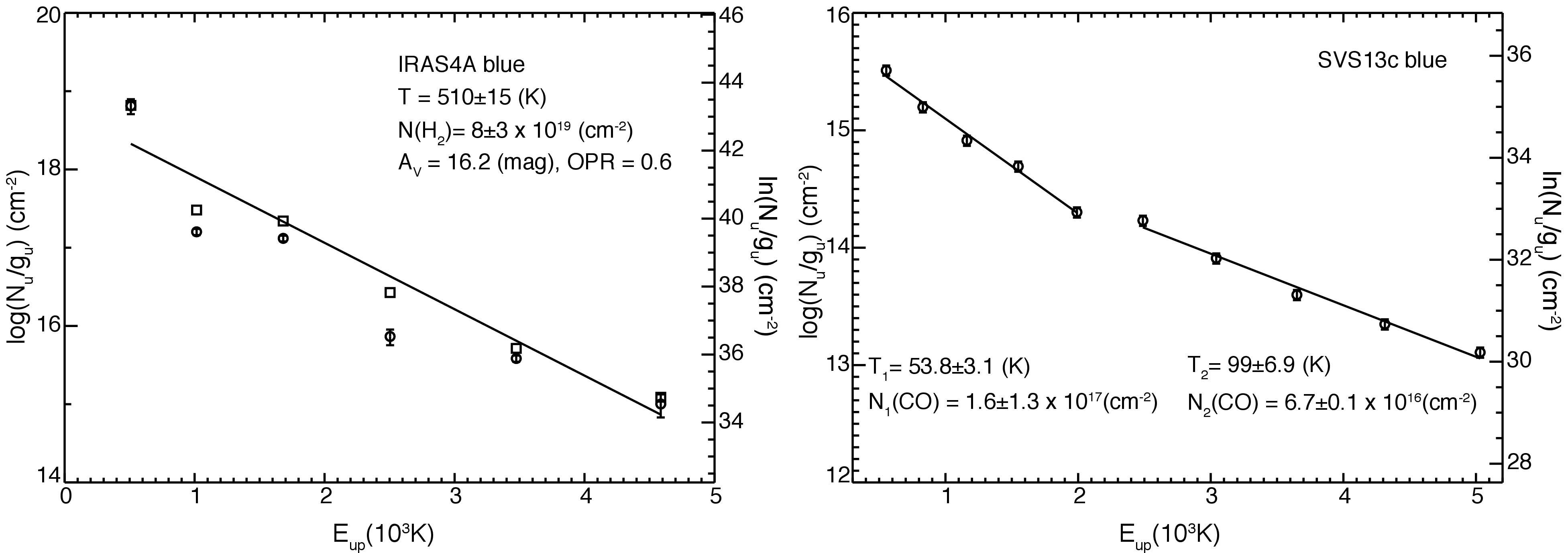}
\caption {(Left:) Excitation diagram for H$_2$; observations (open circles) are fit to calculate the temperature and column density after dereddening (open squares) The distribution of data points can be well approximated with a single temperature component. (Right:) Excitation diagram for CO, where the distribution of data points is best fit with a two distinct temperature components. Excitation diagrams for all outflow regions are presented in Fig.~\ref{onlinefig:3}} 
\label{fig:6a}
\end{figure*}

\begin{figure*}[!h]
\includegraphics[width=17.9cm]{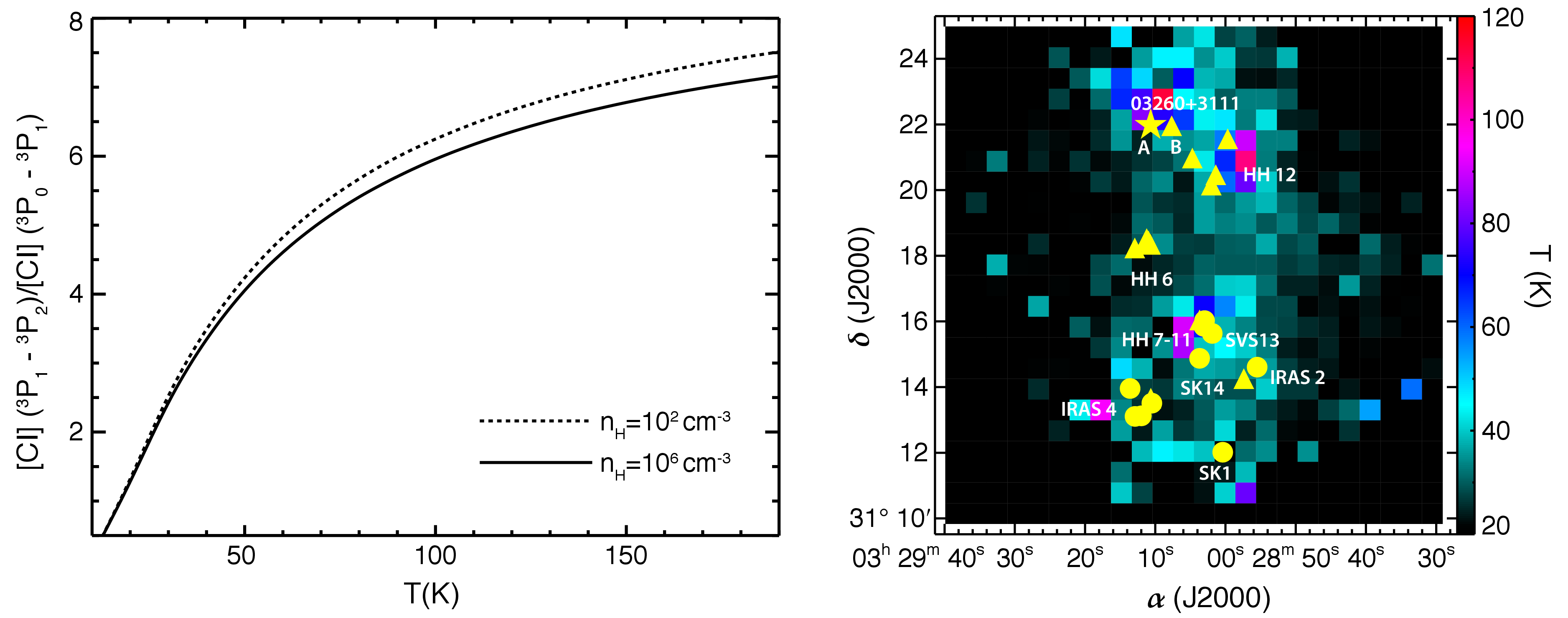}
\caption {(Left:) [CI] line ratio vs temperature diagnostic calculated with RADEX for two hydrogen atom densities. (Right:) Temperature map reconstructed from the [CI] maps presented in Fig.~\ref{fig:4} and the use of the diagnostic.} 
\label{fig:7}
\end{figure*}



\section{Analysis}
 \label{sec:4}

In the following sections we estimate the prevalent excitation conditions responsible for the molecular line emission, and derive the kinematical and dynamical properties  for the gas carried out, as traced by the different outflow tracers. In order to enable comparisons between this work and the results from \citetalias{Dionatos:17a} and \citet{Plunkett:13a}, in the following sections we assume the same driving sources and use the same elliptical outflow loci as defined in \citet{Plunkett:13a}. These are shown for reference in Fig.~\ref{fig:6}, superimposed on the CO (9-8)  line map.



\begin{table*}[!ht]
\small
\caption{Results form the excitation analysis}
\label{tab:1}
\centering
\begin{tabular}{l  c c c c    c    c c c c   }

\hline\hline
Source/outflow  & T(H$_2$)  & N(H$_2$) & A$_V$ & OPR &  &T$_1$(CO) & N$_1$(CO)  & T$_2$(CO) & N$_2$(CO)\\   
lobe  & (K) & (10$^{19}$cm$^{-2}$) & (mag)& &   & (K) & (10$^{16}$cm$^{-2}$) & (K) & (10$^{16}$cm$^{-2}$)   \\
\hline
SVS13A blue  & 520$\pm$20 &  9.2$\pm$0.2 &  17.8 &  1.4 &   &  61.3$\pm$2.0 &  14.0$\pm$5.0 &  93.6$\pm$5.0 &  14.0$\pm$2.0 \\
SVS13A red  & 325$\pm$47 &  7.1$\pm$3.9 &  $\cdots$ &  1.2 &   &  49.4$\pm$1.5 &  16.0$\pm$6.0 &  92.1$\pm$5.3 &  8.6$\pm$0.1 \\

SVS13C blue  & 337$\pm$29 &  6.2$\pm$3.5 &  $\cdots$&  1.1 &   &  53.8$\pm$3.1 &  16.0$\pm$13.0 &  98.9$\pm$6.9 &  6.7$\pm$0.1 \\
SVS13C red & 508$\pm$15 &  7.1$\pm$2.2 &  14.4 &  1.1 &   &  49.6$\pm$1.5 &  27.0$\pm$15.0 &  98.7$\pm$5.5 &  11.0$\pm$1.0 \\

IRAS2A W-E blue   & 295$\pm$30 &  4.5$\pm$3.0 &  $\cdots$ &  1.1 &   &  48.8$\pm$1.6 &  4.3$\pm$2.8 &  100$\pm$8.0 &  1.5$\pm$0.3 \\
IRAS2A W-E red  & 475$\pm$50 &  5.4$\pm$2.4 &  16.9 &  1.1 &   &  57.7$\pm$4.0 & 3.9$\pm$0.3  &  117.1$\pm$9.5 &  1.7$\pm$0.2 \\

IRAS2A S-N blue & 520$\pm$45 &  8.6$\pm$6.4 &  15.0 &  0.5 &   &  46.1$\pm$2.5 &  8.1$\pm$0.7 &  111.9$\pm$7.7 &  3.8$\pm$0.2 \\
IRAS2A S-N red & 530$\pm$10 &  3.9$\pm$2.1 &  14.8 &  0.7 &   &  48.0$\pm$1.9 &  13.0$\pm$8.0 &  96$\pm$6.8 &  3.8$\pm$0.1 \\

IRAS4A blue & 510$\pm$15 &  8.0$\pm$3.0 &  16.2 &  0.6 &   &  59.2$\pm$3.0 &  3.3$\pm$2.0 &  121.7$\pm$6.6 &  3.5$\pm$0.3 \\
IRAS4A red& 535$\pm$30 &  2.4$\pm$1.2 &  9.4 &  0.6 &   &  68.7$\pm$5.7 &  5.2$\pm$0.4 &  124$\pm$5.4 &  6.1$\pm$0.4 \\

SK14 blue & 455$\pm$35 &  5.0$\pm$3.2 &  18.7 &  0.7 &   &  51.9$\pm$3.0 &  4.3$\pm$0.3 &  75.7$\pm$4.8 &  2.2$\pm$0.4 \\
SK14 red  & 239$\pm$32 &  3.8$\pm$1.2 &  $\cdots$ &  0.5 &   &  64.9$\pm$3.2 &  4.5$\pm$0.4 &  107.7$\pm$5.1 &  3.0$\pm$0.3 \\

SK1 blue & 445$\pm$57 &  4.2$\pm$2.7 &  13 &  0.6 &   &  33.7$\pm$1.7 &  3.8$\pm$0.5 &  80.9$\pm$11.8 &  1.8$\pm$0.7 \\
SK1 red   & 517$\pm$61 &  2.5 &  10 &  0.5 &   &  39.5$\pm$2.6 &  5.2$\pm$0.8&  133.4$\pm$23.3 &  7.3$\pm$0.2 \\

\hline

\end{tabular}
\end{table*}

\subsection{Excitation conditions of H$_2$ and CO} \label{sec:4.1}

The physical conditions driving the excitation of excited gas can be directly constrained by means of excitation diagrams. These diagrams display the logarithm of the column density over the statistical weight as a function of the upper level energy for a range of transitions of a species. Assuming that the gas is thermalized and the medium is optically thin, a straight line can be fitted to excitation diagram points, thus determining the rotational temperature (T$_{rot}$) of the gas from the inverse slope and the total column density from the intercept \citep[e.g.,][]{Dionatos:13a, Yang:18a}.  
In the case of molecular hydrogen, we can also estimate the ortho-to-para ratio (OPR) and the optical depth using deviations from the assumed linear distribution of data points on the excitation diagrams. In particular, deviations of the ortho-to-para ratio from the equilibrium value of 3 will appear as vertical displacements between the ortho and para transitions forming a ``saw-tooth'' pattern. The ortho-to-para ratio can be best determined by examining the alignment of the S(5) with the neighboring S(4) and S(6) rotational H$_2$ transitions \citep{Dionatos:10a}, however due to the limited sensitivity of the IRS maps at shorter wavelengths we compare the alignment of the S(1) to the S(0) and S(2) transitions. In addition, the H$_2$ S(3) transition at 9.7 $\mu$m is sensitive to the amount of dust along the line of sight, being located within a wide-band silicate absorption feature at the same wavelength; consequently, a visual extinction value can be estimated by comparing the displacement of this line with in relation to the S(2) and S(4) transitions \citep[e.g.][]{Dionatos:09a}. Using these extinction estimations we deredden the H$_2$ lines using the extinction law of \citet{Chapman:09a} before deriving any measurements from the excitation diagrams.  

An example of an  H$_2$  excitation diagram is provided in Fig.~\ref{fig:6a} while diagrams for all outflow lobes are provided in Appendix~\ref{app:A}. Excitation temperatures, column densities, ortho-to-para ratio and visual extinction values derived from the excitation analysis are summarized in Table~\ref{tab:1}. As a general trend the H$_2$ emissions  can be well approximated with a single temperature component (see also Fig.~\ref{onlinefig:3}). The H$_2$ S(0) line intensity appears to be overestimated in most of the cases, most likely due to residual emission from a persisting rogue pixel in the IRS scans. This results in underestimating the H$_2$ ortho-to-para ratio (see Table~\ref{tab:1}) and therefore derived ortho-to-para ratios should be considered only for comparisons between the different regions. Overestimating the S(0) intensity levels however does not appear to compromise the derivation of excitation temperature or column density as it does not have a significant impact on the quality of linear fits, with a couple of exceptions where very few H$_2$ transitions are detected. For those cases the excitation temperature is slightly underestimated and column density overestimated. The derived H$_2$ excitation temperatures lie between $\sim$ 400 and 500\,K and column densities between 10$^{19}$ and 10$^{20}$~cm$^{-2}$. Visual extinction ranges between $\sim$ 10 and 20 mag, in good agreement with extinction estimations from near-IR and X-ray observations. \citep{Walawender:08a}.

A typical excitation diagram for carbon monoxide is shown in Fig.~\ref{fig:6a}; more diagrams for all other lobes are included in Appendix~\ref{app:A}. The distribution of CO lines can be well approximated with two linear segments, corresponding to two temperature components.  These correspond to lines detected with the SSW and the SLW SPIRE modules and, in some cases, appear to have a small offset. While this can result from a miscalibration between the two modules, the offset does not appear to be consistent for all lobes. CO lines from $J_{up}$~9  upwards can be optically thin as it is shown for a large number of embedded sources in \citet{Yildiz:12a} and \citet{Yang:18a}. Indeed, optical depth affects mostly SLW lines (as discussed in more detail in the previous section) and as a result, the derived number of molecules in the upper state level of each transition for  $J_{up}$<9 is underestimated, creating the apparent mismatch in the excitation diagrams. Therefore column density effects may result in slightly underestimating the total column density of the cold component, however derived temperature should remain unaffected. Column densities range between $\sim$ 10$^{16}$ and 10$^{17}$ cm$^{-2}$ and differ by a factor of $\sim$2-3 between the cold and the warm component, while temperatures range between $\sim$50 and $\sim$100\,K for the two components. The results from the excitation analysis are detailed for each molecule, source and outflow lobe in Table~\ref{tab:1}.

\subsection{Excitation of $[\ion{C}{i}]$ }

Given the quite different morphology revealed in the maps of the two  $[\ion{C}{i}]$ lines detected with SPIRE, we attempt to determine the possible underlying excitation mechanisms. We employ the ratio of the $^3$P$_2$ - $^3$P$_1$ transition at  370\,$\micron$  over the $^3$P$_1$ - $^3$P$_0$ transition at  610\,$\micron$ as a temperature probe, since these lines arise from levels that are separated well in energies and therefore their population depends mainly on the efficiency of the excitation process. We analyzed the $[\ion{C}{i}]$ observations employing RADEX, which is a radiative transfer code based on the escape probability approximation \citep{vanderTak:07a}. Collisional rates for $[\ion{C}{i}]$ \citep{Schroder:91a} were taken from the LAMDA\footnote{https://home.strw.leidenuniv.nl/~moldata/} database. We ran a grid of RADEX models for a range of densities $n$ between 10$^2$ and 10$^6$ cm$^{-3}$,  kinetic temperatures between 10 and 200\,K and total carbon column densities ranging from 10$^{15}$ to 10$^{17}$\,cm$^{-2}$. The $[\ion{C}{i}]$ line ratio appears insensitive to variations of the total density as shown in the left panel of Fig.~\ref{fig:7} and depends mostly on the temperature. Using this diagnostic, we attribute a single approximate temperature to the line ratio corresponding in each SPIRE spaxel to create the temperature map shown in the right hand panel of Fig.~\ref{fig:7}.  $[\ion{C}{i}]$ appears to be warmer at the base of outflows close to known protostars reaching temperatures of $\sim$ 100\,K, while in more diffuse regions appears to trace cold gas at 30 - 50\,K. 


\begin{table*}[!ht]
\small
\caption{ CO, H$_2$ and H$_2$O mass flux, jet dynamical timescales.}
\label{tab:2}
\centering
\begin{tabular}{l  c c c c c c   c   c c c c c c   }

\hline\hline
Source  &  \multicolumn{6}{c}{Blue Lobe} &  &\multicolumn{6}{c}{Red Lobe}  \\ \cline{2-7} \cline{9-14 } 
  & $a^a $ & $v_t^b$& t$_{dyn}$ &$\dot{M}_{H_2}$ & $\dot{M}_{CO}$ & $\dot{M}_{H_2O}$  &   & $a^a$ & $v_t^b$& t$_{dyn}$ &$\dot{M}_{H_2}$ & $\dot{M}_{CO}$ & $\dot{M}_{H_2O}$   \\ \cline{5-7} \cline{12-14 } 
  & ($\arcsec$) & (km s$^{-1}$) & (10$^3$ yr)& \multicolumn{3}{c}{ (10$^{-6}$ M$_{\odot}/yr$)} &   & ($\arcsec$) & (km s$^{-1}$) & (10$^3$ yr)& \multicolumn{3}{c}{ (10$^{-6}$ M$_{\odot}/yr$)}  \\
\hline

SVS13A & 122 & 20 & 6.8  & 3.1  & 23.6	  &0.16  	& &		 163 &  20 & 9.1& 1.7 & 18.7 &0.07  \\
SVS13C & 212 & 100 & 2.4 & 12.7 & 14.1 &0.82	 	& &		 180  &  10 & 20 &1.25 & 23.8  &0.08 \\
IRAS2A W-E & 108 & 15 & 8.1& 0.29 & 1.39	 &0.01	& &    84 &  15 & 6.2 & 0.36 & 1.32 &0.02 \\
IRAS2A S-N & 188 & 50 & 4.1  & 3.8 & 29.1	&0.19 	& &	  133 &  50 & 2.9 & 3.6 & 60.1 &0.32 \\
IRAS4A & 104 & 100 & 1.2  & 3.9 & 8.1	&0.18 	& &		 125 &  100 & 1.4 & 1.43 & 15.5 &0.33 \\
SK14 & 71 & 30  & 2.6 & 1.2 & 5.2 &0.02	 	& &		 	58  &  30 & 2.2 & 0.8 & 4.9 &0.03 \\
SK1 & 165 & 80 & 2.3 & 1.8 & 10.2	&0.03 	& &		     165 & 80 & 2.3 & 2.2 & 13.6 &0.08 \\

\hline

\end{tabular}
\tablefoot{$^a $  Outflow lengths are adopted from \citep{Plunkett:13a}.\\
 $^{b} $Tangential velocities are adopted from \citet{Raga:13a}. }
\end{table*}

\begin{table*}[!ht]
\small
\caption{Mass flux estimations for  jets and outflows in NGC\,1333}
\label{tab:3}
\centering
\begin{tabular}{l  c c c    c c c c c  c  }

\hline\hline
Source & T$_{bol}^a$ & L$_{bol}^a$ & t$_{dyn}^b$   & $\dot{M}_{H_2}$  & $\dot{M}_{CO}$ & $\dot{M}_{H_2O}$ & $\dot{M}_{[\ion{O}{i}]}^c$ & $\dot{M}_{[\ion{O}{i}] - shock}^c$  \\ \cline{5-9}
  & (K) & L$_{\odot}$ & (10$^3$ yr)&       \multicolumn{5}{c}{ (10$^{-6}$ M$_{\odot}/yr$)}   \\
\hline

SVS13A & 250 & 59 & 8.0  &   			4.8	    & 42.3 &0.23 &0.19 & 3.81  \\   
SVS13C & 36 & 4.9 & 11.2 &  			13.9    & 37.9&0.90   &0.18 & 3.34  \\
IRAS2A W-E & 57 & 76 & 7.1 &    		0.65     & 2.71&0.03  & 0.01 & 0.12  \\
IRAS2A S-N & 57 & 76 & 3.5 & 			 7.4   & 89.2&0.51  & 0.08 & 0.76  \\
IRAS4A & 43 & 5.8 & 1.3 & 		 	5.3	     &  23.6&0.51  & 0.14 & 0.49  \\
SK14 & 59 & 0.2 & 2.4 & 		 		2.0 	  & 10.1&0.05  & 0.09 & 0.53 \\
SK1 & 32 & 0.7 & 2.3 & 		 		4.0       & 23.8&0.11  & 0.03 & 0.18  \\

\hline

\end{tabular}
\tablefoot{$^a $ Source bolometric temperatures and luminosities are from \citet{Plunkett:13a}. \\
 $^{b} $ Dynamical scale is the average from the two lobes, as reported in Table~\ref{tab:2}. \\
 $^{c} $ [OI] mass flux values are from \citetalias{Dionatos:17a} and is reported here to enable comparisons.}
\end{table*}

\subsection{H$_2$, CO and H$_2$O mass flux carried by outflows}

Being the most abundant molecule, the observations of H$_2$ provide a unique opportunity to directly derive the total mass flux carried out by the outflows acting in NGC\,1333. We employ the method first described in \citet{Dionatos:09a} where the mass flux is derived from the H$_2$ column density through the following relation:
\begin{equation}
\frac{dM_{H_2}}{dt}= 2\mu m_H \times N(H_2)\times A \times \frac{du_t}{dl_t}
\label{eq:1}
\end{equation}
where $\mu$ is the mean atomic weight, m$_H$ is the mass of the hydrogen atom, $N(H_2)$ is the molecular hydrogen column density, $A$ corresponds to the sampled area, $du_t$ is the tangential velocity and $dl_t$ corresponds to the assumed length of the outflow. The factor 2 in eq.~\ref{eq:1} accounts for the two hydrogen atoms comprising a hydrogen molecule. The area A and length $dl_t$ of the outflow correspond to the ellipses circumscribing the outflows in NGC\,1333 as described in \citet{Plunkett:13a}. The column density for these areas is calculated in the excitation diagrams of the previous section. Outflow tangential velocities are taken from \citet{Raga:13a}, based on proper motion studies of mid-infrared knots traced in multi-epoch Spitzer/IRAC images. Knots emission in IRAC bands is dominated by H$_2$ emission lines tracing shocks and therefore provide reliable measurement of the same gas recorded in the IRS spectro-images presented here. 

Eq.~\ref{eq:1} can be modified to provide the mass flux carried out by any molecule $X$ from its column density $N(X)$ subtended in an area $A$ of length $dl_t$ along the path of motion, for gas moving with a velocity as measured on the plane of the sky $du_t$: 

\begin{equation}
\frac{dM_X}{dt}= \mu m_H \times N(X)\times A \times \frac{du_t}{dl_t}
\label{eq:2}
\end{equation}

In the case of CO, column densities were calculated by means of the excitation diagrams in the previous section. From the two temperature components we employ the measurements corresponding to the warmer one ($\sim$100\,K) as it suffers less from self-absorption and can therefore record more reliably the bulk of the outflow gas. In order to account for the total amount of gas we employed a ``canonical'' CO abundance value of $10^{-4}$. The physical conditions dominating  H$_2$O excitation cannot be determined from a single transition, as is the case with the data available in this study. However an estimation of the water column density can be derived using the relation described in \citet{Tafalla:13a}, which is based on detailed local velocity gradient (LVG) model solutions applied to a number of different protostellar outflows:

\begin{equation}
N\mathrm{(H_2O)\,  [cm^{-2}]}= 1.2 \times 10^{12} I \mathrm{\, (557~GHz)\, [K\, km\, s^{-1}],}
\label{eq:3}
\end{equation}
where $I$ is the water line intensity of the water transition at 557~GHz. In addition, we employed an abundance ratio for water-ice of $5\times10^{-5}$ \citep{vanDishoeck:14a}, assuming that all water is efficiently released into the gas phase even at moderate outflow velocities \citep{Suutarinen:14a}. Lacking any proper motion observations for either CO or H$_2$O, we employ the tangential velocities from \citet{Raga:13a} that were used also in mass flux estimations of H$_2$. It is worth mentioning here that the same tangential velocities were also used for the mass flux calculations of the [\ion{O}{i}] jet in \citetalias{Dionatos:17a}. The assumption of a common velocity for all species is a simplification that introduces uncertainties in our calculations, albeit it enables direct comparisons. Results from the mass flux calculations for H$_2$, CO and H$_2$O are reported in Table~\ref{tab:2} separately for each outflow lobe, along with the length for each outflow, the tangential velocity employed and the corresponding dynamical timescale that is calculated from the relation:

\begin{equation}
t_{dyn}=\frac{dl_t}{du_t}
\label{eq:4}
\end{equation}

In Table~\ref{tab:3} we report the total mass flux corresponding to each protostar for all molecules discussed here, along with the source properties such as the bolometric temperature and luminosity from \citet{Plunkett:13a}. Dynamical timescales in Table~\ref{tab:4} are the average for the two lobes reported in Table~\ref{tab:3}. To enable direct comparisons, in the last two columns of Table~\ref{tab:4} we report the mass flux corresponding to the [\ion{O}{i}] jet estimated using two different methods in \citetalias{Dionatos:17a}.  

As a general trend, the CO mass flux is a factor of $\sim$10 higher compared to the one corresponding to H$_2$ which in its turn is another factor of  $\sim$10 higher when compared to the mass flux of H$_2$O. The water mass flux is of the same order of magnitude as the one corresponding to the  [\ion{O}{i}] emission.

\begin{table*}[!ht]
\small
\caption{ Momentum and energy estimations of jets and outflows in NGC\,1333. }
\label{tab:4}
\centering
\begin{tabular}{l  c c c c c  c  c c c c c   }

\hline\hline
Source & T$_{bol}^a$ & L$_{bol}^a$ & t$_{dyn}^b$ & P$_{H_2O}  ({\times 10^{-2}})$ & P$_{[\ion{O}{i}]}^c ({\times 10^{-2}})$ & P$_{CO}^d$ & &E$_{H_2O}({\times 10^{42}})$ & E$_{[\ion{O}{i}]}^c({\times 10^{43}})$  & E$_{CO}^d({\times 10^{43}})$ \\ \cline{5-7}  \cline{9-11}
  & (K) & L$_{\odot}$ & (10$^3$ yr)&  \multicolumn{3}{c}{(M$_{\odot}$ km s$^{-1}$)} & & \multicolumn{3}{c}{(erg)}  \\
\hline

SVS13A & 250 & 59 & 8.0 & 2.0  &  8.2 & 4.6	 	& & 7.7 & 	17.7	    & 17.6  \\ 
SVS13C & 36 & 4.9 & 11.2 & 4.0 & 6.8  & 5.0 	 & 	& 14.9 &	14.3    & 21.8   \\
IRAS2A W-E & 57 & 76 & 7.1& 0.1 &  0.3 & 0.6	 & 	 & 0.4 & 1.0     & 2.1 &   \\
IRAS2A S-N & 57 & 76 & 3.5& 1.8 & 1.7 & 3.7	 & 	 & 6.4 &	 4.8   & 12.8   \\
IRAS4A & 43 & 5.8 & 1.3& 0.7 & 1.1 & 1.7	 	& 	& 2.3 & 	2.1	     & 3.7   \\
SK14 & 59 & 0.2 & 2.4& 0.1 & 0.9 & 0.7 		& 	& 0.5& 		1.6 	  & 1.8  \\
SK1 & 32 & 0.7 & 2.3& 0.3 & 0.5 & 1.0	 	& 		& 1.1& 		1.6       & 3.1   \\

\hline

\end{tabular}
\tablefoot{$^a $ Source bolometric temperatures and luminosities along with the CO outflow momenta and energies are from \citet{Plunkett:13a} \\
 $^{b} $ Dynamical scale is the average from the two lobes, as reported in Table~\ref{tab:2}. \\
 $^{c} $ [OI] outflow momenta and energies are from \citetalias{Dionatos:17a}.\\
 $^{d} $ CO outflow momenta and energies are from \citet{Plunkett:13a} and correspond to the CO $J=$2-1 line.}
\end{table*}

\subsection{H$_2$O momentum and energy}


Water lines observed with HIFI are resolved and therefore the momentum carried out by the water excited in outflows can be calculated as discussed in \citetalias{Dionatos:17a} using the following relation:

\begin{equation}
P_{H_2O} = \sum_{v_{bin}} M_{v_{bin}}  \lvert v - v_{r} \rvert
\label{eq:5}
\end{equation}
\noindent
where the sum acts over the mass per velocity bin, $v$ is the mean velocity for each velocity bin and $v_{r}$ is the rest velocity for the water molecule. The mass per velocity bin ($M_{v_{bin}}$) is derived from the column density (eq.~\ref{eq:3}) but in this case calculations are applied per velocity bin.

\begin{equation}
M_{v_{bin}}= \mu m_H \times N_{v_{bin}}\times A 
\label{eq:6}
\end{equation}

The outflow energy traced by the H$_2$O  line can be derived according to the following relation:

\begin{equation}
E_{H_2O} = \frac{1}{2}\sum_{v_{bin}} M_{v_{bin}}  \lvert v - v_{r} \rvert^{2}
\label{eq:7}
\end{equation}
\noindent
where the mass per velocity bin is defined in equation~\ref{eq:6} and velocities follow the same notation as for the momentum. The uncertainties in the mass derivation discussed above also affect the estimation of the momentum and energy deposited by the outflows.  Water momentum and energy values for all outflows are reported in in Table~\ref{tab:4} along with the embedded source properties. In the same table we provide the corresponding momentum and energy values for the [\ion{O}{i}] jet and the CO outflow (from \citetalias{Dionatos:17a} and \citet{Plunkett:13a}, respectively), for comparisons. The momentum traced by water is in good agreement with the values derived for [\ion{O}{i}] but much lower than the momentum derived for CO. In addition water appears to trace a less energetic outflow component compared to  CO or [\ion{O}{i}].

\subsection{H$_2$O abundance}

Water can be formed and excited through a multitude of diverse pathways. A direct determination of the water abundance requires comparison to another molecule of known abundance which is excited through the same mechanism and tracing similar conditions to those traced by water. Following these lines, we calculated the water abundance based on comparisons between the H$_2$O and H$_2$ column densities measured in individual outflows based on the observation that the two molecules are co-spatial and appear to have similar line strengths \citep{Tafalla:13a}. We also derived water abundances comparing the column density of H$_2$O to that of the warm component traced by CO, based on the observation that intermediate J CO and water line profiles appear to closely correlate and therefore trace the same volume of gas \citep{Kristensen:17a}. To convert CO-based water abundances to H$_2$ we assumed a nominal CO abundance of 10$^{-4}$, however direct estimations in outflows indicate that this number can be as low as 10$^{-5}$ \citep{Dionatos:13a, Dionatos:18a}. We report the results of these comparisons for the outflows driven by each source in Tab.~\ref{tab:5}. H$_2$O abundances based on comparisons to H$_2$ average to $\sim5.3\times 10^{-6}$ while those based on comparisons to warm CO average to $\sim7\times 10^{-7}$.   

\subsection{Molecular abundances around 03260+3111}

In addition to outflow regions we measured column densities for all molecules examined in this work within a circular region extending to a 30$\arcsec$ radius around the B-star 03260+3111 to the NE. Excitation of gas in that region is dominated by energetic photons as also evidenced from the detection of very bright [\ion{O}{I}] and [\ion{C}{II}] emission presented in \citetalias{Dionatos:17a} \citep{Hollenbach:97a}, and can therefore be used as a reference to compare with the excitation of gas occurring along outflows. Using the same line of arguments as in the previous paragraph,  we derive a water abundance of $4.6\times10^{-8}$ based on comparisons with CO (Tab.~\ref{tab:5}). The abundance based on comparisons with molecular hydrogen is another factor of $\sim6$ lower than the CO-derived ones and 10$^3$ times lower when compared to H$_2$-based abundances in outflows. This large discrepancy can be explained by a number of parallel-acting physical mechanisms.  Photodissociation of molecular hydrogen may occur with absorption of a photon in the Lyman and Werner bands  while at moderate column densities FUV lines become optically thick and H$_2$  begins to self-shield against radiation at $A_V\simeq2$ \citep{Hollenbach:97a}. Water in contrast absorbs UV radiation less selectively than H$_2$ and shields less efficiently, so it can be dissociated at higher column densities even for moderate radiation field strengths. Furthermore, when H$_2$ is dissociated then the reaction H$_2$O + H $\rightarrow$ OH + H$_2$ will further collisionally dissociate water replenishing at the same time the H$_2$ reservoir \citep{Kristensen:17a}. Similar to water, CO shields less efficiently and at higher densities compared to H$_2$ which can explain the smaller discrepancy of the CO-based water abundance measured around 03260+3111 when compared to outflows.   

Interestingly, when comparing CO and H$_2$ column densities we derive a CO abundance of 1.5$\times 10^{-5}$. This measurement provides additional evidence that lower CO abundances than the ``canonical'' value of 10$^{-4}$ may also apply in conditions very different than the ones prevailing in outflows, and hints that a lower value may be more appropriate in dense regions independently of the particular excitation conditions. 

\begin{table}[!ht]
\small
\caption{Water abundance in outflows}
\label{tab:5}
\centering
\begin{tabular}{l  c c  }

\hline\hline
Source &  X$_{H_2O}^a  ({\times 10^{-6}})$ & X$_{H_2O}^b ({\times 10^{-7}})$  \\ 
\hline

SVS13A & 4.7 & 3.3   \\ 
SVS13C & 6.4 & 5.1    \\
IRAS2A W-E & 4.6 & 8.2    \\
IRAS2A S-N & 6.6 & 7.8    \\
IRAS4A & 9.6 & 9.9    \\
SK14 & 2.5 & 12.1   \\
SK1 & 2.7 & 2.2   \\
03260+3111$^c$ & 0.007 & 0.46\\
IRAS4A$^d$ & $\cdots$ & 17.6    \\
\hline

\end{tabular}
\tablefoot{$^a $Based on comparison between water and molecular hydrogen column densities. 
 $^{b} $Based on comparisons between the water and ``warm'' carbon monoxide column densities and adopting a CO abundance of 10$^{-4}$. 
$^{c} $Photon-dominated region.
$^{d} $On-source} 
\end{table}

\section{Discussion}
\label{sec:5}


In the following discussion we attempt to shed light on physical processes responsible for the excitation the various molecules and the different kinematical/dynamical components they trace. Our aim is to set a reference for the use and limitations of different tracers in constraining jet and outflow properties.  Water emission in particular is discussed in the context of its abundance variations along with possible formation and excitation pathways.

\subsection{H$_2$ and CO}\label{sec:5.1}

Molecular hydrogen appears brightest around the intermediate-mass sources to the NE of the Spitzer maps indicating heating under the influence of energetic UV radiation from those sources. In particular the H$_2$ emission morphology around BD+30459 traces a shell-like structure with the source positioned at the center. In the central regions of the map H$_2$ is excited around HH6, HH12 and SVS13; It appears brightest at the location of the HH 7-11 outflow and also at a symmetrical outflow structure to the northwest of the driving source SVS13. To the south, H$_2$ emission delineates a web of interlaced linear and bow-shaped features tracing extended emission from molecular jets and outflows. Despite the limited sensitivity of the Spitzer maps it is clear that H$_2$ emission does not trace the bulk of the outflowing material but is rather confined in certain regions. This is consistent with the result from \citetalias{Dionatos:17a} that [\ion{O}{i}] maps closely follow the H$_2$ morphology, which indicates that both tracers are excited in shocks. Inline with our analysis, H$_2$  traces warm gas with temperatures of $\sim$450\,K while in more sensitive maps a hot component at T$\sim$1000\,K is also commonly observed \citep[e.g.][]{Maret:09a, Dionatos:09a, Dionatos:10a}. Such an interpretation is also consistent with the intrinsic properties of H$_2$, that, as a very light molecule, has its first excited rotational levels at energies above 300\,K.   

The mass flux traced by H$_2$ is at least an order of magnitude higher compared to that derived from [\ion{O}{i}], under the assumption that the two elements move at the same velocity. When such shock interfaces are examined in detail, H$_2$ appears to be excited at the flanks of a bow-shock while atomic lines become bright at the shock-head \citep{Dionatos:14a} and therefore the assumption of a common velocity between the two species appears to be an oversimplification. If the mass flux traced by the two components is preserved, then [\ion{O}{i}] should be moving a factor between 10 and 20 faster than H$_2$. The mass flux traced by H$_2$ therefore corresponds to the mass of the gas excited in energetic shocks (internal working surfaces) along the jet-propagation path. 

Depending on the excitation conditions $^{12}$CO may trace diverse excitation processes. While low-$J$ transitions are in general optically thick, emission from the line wings can trace high velocity gas in outflows. At the limit emission becomes optically thin \citep[for $J_{up}\sim$ 8-9][]{Yang:18a},  $^{12}$CO appears to trace gas excited in shocks or gas heated under the influence of ultraviolet radiation. This transition is readily evidenced in the CO maps of Fig.~\ref{fig:4}, especially for the transitions falling in the short-wavelength SPIRE module.  

Comparing the mass flux derived from the CO lines mapped by SPIRE to the H$_2$ mass flux, the latter is found lower by an order of magnitude on average.  When the abundances of the two molecules are considered this result appears counterintuitive and can only be explained in the context of the H$_2$ excitation in shocks.  In this respect, low-excitation CO lines pick up the bulk of the outflowing material in a turbulent layer surrounding the protostellar jet. H$_2$ in contrast, despite being the most abundant molecule does not capture the total mass of the gas entrained in outflows. It is only CO lines of higher excitation energy (above $J_{up} \sim14$) that start tracing gas excited in shocks rather than the turbulent mixing layer and therefore it is only those, higher-$J$ lines that can trace the same excitation conditions as H$_2$ \citep[e.g.,][]{Yildiz:13a, Kristensen:17a, Yang:18a}. 




\subsection{H$_2$O}


Continuing the discussion from Section \ref{sec:3.2}, the water-line profiles can be decomposed into two distinct components: a narrow, weaker component moving close to the systemic velocity of the cloud with a width of just $\sim$2--3~km~s$^{-1}$, and a faster component forming line wings which extend to $\sim$25~km~s$^{-1}$ and is associated to outflows. The distribution of the narrow component in Fig.~\ref{fig:3}, shows that it strongly peaks around the intermediate mass sources to the north and especially around the B-star 03260+3111. Strong emission from [\ion{O}{i}], [\ion{C}{ii}], H$_2$, [\ion{C}{i}] and CO at the same location provide evidence that UV radiation from the star creates a photon dominated region (PDR) in its surroundings \citep{Hollenbach:97a}. The narrow emission component is therefore associated with the influence of UV radiation and when traced, it can help diversify between the underlying excitation processes. 
Other regions in the narrow component map (Fig.~\ref{fig:3}) where water emission can attest the influence of UV heating are found to the west of HH~12, along HH~7-11 and in the vicinity of the IRAS4 system. Maps of $^{12}$CO transitions with J$_{up}\geq9$ display emission maxima at exactly the same locations, providing evidence that intermediate excitation CO lines can have an important component that is driven by UV heating. These findings are also in agreement with the analysis of  \citet{Yildiz:12a} where the narrow component of the CO 6-5 line along the outflow cavities from IRAS4A/4B is interpreted to be resulting from the UV heating of the gas. When comparing the excitation conditions and outflow kinematical properties, no apparent difference is found between outflows with and without signatures of UV heating indicating that the influence of UV radiation is rather localized and smoothed out at the outflow scales used for our calculations. This is also consistent with the distribution of  [\ion{C}{ii}] which appears prominent in dense photodissociation regions but is not detected along outflows. 

The morphology of the faster H$_2$O component is very similar to the emission maps of jets and outflows traced by [\ion{O}{i}], CO and H$_2$.  The bulk of the H$_2$O outflow velocity correlates well with the velocity distribution for the bulk of [\ion{O}{i}], with a major difference being that H$_2$O emission traces velocities up to 15~km~s$^{-1}$ as opposed to 50~km~s$^{-1}$ reached in the [\ion{O}{i}] maps. At those velocities we expect that most of the water should be released into the gas phase through sputtering \citep{Suutarinen:14a, Kristensen:17a}. The distribution of the integrated H$_2$O intensity maps does not appear to follow the bulk H$_2$O outflow velocity pattern (see Fig.~\ref{fig:3}), indicating that water emission is deficient in fast shocks and predominantly originates from material coasting in outflows after having been accelerated in shocks. Line wings of water trace velocities up to $\sim$30~km~s$^{-1}$, which is in agreement with the predictions for the dissociation of the molecule at higher velocities in J-shocks \citep{Suutarinen:14a}. Water reformation through warm gas-phase reactions is possible in the post-shock cooling zone, however estimated timescales are very long for typical outflow densities \citep{Kristensen:17a}. 

Based on pointed HIFI observations of water lines around a sample of embedded protostars, \citet{Mottram:14a} proposed a decomposition of the line profile to three components: (i) a very narrow component close to the cloud systemic velocity corresponding to emission from the envelope, (ii) a intermediate component arising from shocks along the cavity walls and (iii) fast components corresponding to molecular bullets or all other locations where the low-velocity wind directly impacts the envelope. The narrow component map in Fig.~\ref{fig:3} shows that energetic radiation can escape inside outflow cavities but it can also be generated in-situ from the action of fast shocks (e.g. HH\,7-11). The maximum velocity reached for the bulk of the H$_2$O gas in the line-centroid maps is $\pm$~15 km s$^{-1}$, which reflects exactly the same limits set for the ``cavity shocks'' in \citet{Mottram:14a} and possibly traces gas that is coasting in outflows after being accelerated. These boundaries also reflect a possible a physical limit where most of water is dissociated in shocks and not replenished at the same rate in the post-shock cooling zone. Water emission at higher velocities is rather localized as demonstrated in the velocity-channel maps of Fig.~\ref{onlinefig:1}, which is consistent with the formation/excitation in molecular bullets.  
 
Water emission peaks in the surroundings of the IRAS\,4 system as seen in the integrated emission maps of Fig.~\ref{fig:3}, while the maximum is found at the location of IRAS\,4A. Higher excitation water emission lines in the outflows of IRAS\,4B have been previously reported by \citet{Herczeg:12a} where a $\sim$50\% of the cooling was measured to take place through water lines \citep{Karska:18a}. The water line profile around IRAS\,4A (Fig.~\ref{fig:2}) displays signatures of both radiative and outflow components. High velocity outflows are also prominent in the water velocity channel maps of Fig.~\ref{onlinefig:1}. Signatures of UV-heated gas extending in the outflow cavities of IRAS\,4A have been isolated in the narrow component of $^{12}$CO and $^{13}$CO 6-5 lines \citep{Yildiz:12a}. Such radiation can alternatively be produced in accretion shocks very close to the forming star and then scatter to large distances in the outflow cavities before being absorbed heating up the material in the cavity walls \citep{Spaans:95a}. 

The water abundance with respect to CO for a circular region of a 10$\arcsec$ radius centered on IRAS\,4A measures 1.7$\times 10^{-6}$, which is a factor of 2 higher than the abundance measured in the outflows of the same source and at least 50\% higher than any other measurement in outflow positions. The angular resolution of the CO and H$_2$O maps does not allow us to isolate the source from the outflow positions, however the observed abundance enhancement must be resulting from the area very close to IRAS\,4A which is not included in the abundance calculations of the outflows. These findings are consistent with the best-fitting water abundance profile used to model the profiles of multiple water transitions around IRAS\,4A \citep{Mottram:13a}, where the abundance is found to sharply increase in the inner 100 au from the source. Energetic radiation from the central source cannot account for the observed water enhancement as the example of  03260+3111 shows that the net effect of UV-heating would be to decrease rather than increase the water abundance. Gas phase chemistry in outflows could increase the water abundance up to a factor of 2 \citep{Suutarinen:14a}, however it is hard to explain why such an enhancement would selectively take place at the base of outflows and not further out. Yet the outflows of IRAS\,4A appear to be an exceptional case; following the Gaussian decomposition of \citet{Mottram:14a}, these outflows possess a uniquely fast ``spot-shock'' component which is however not shifted from the source velocity and therefore appears to arise from the envelope cavities rather than a fast jet.

In this respect, a water reservoir may preexist frozen onto the dust grains in the protostellar envelope, and is evaporated where temperatures reach above $\sim$100\,K. In such a case similar water reservoirs should exist in the envelope of most, if not all YSOs, however no similar behavior is traced anywhere else in NGC\,1333. This suggests that IRAS\,4A is a particular case unlike any other YSO in our maps. Indeed, when observed with interferometers IRAS4A is resolved into a binary system, where the weaker (A2) component is displaying complex chemical signatures of a hot corino \citep{Bottinelli:04a, Sahu:19a}. Interferometric observations of H$_2^{18}$O \citep{Persson:12a} confirm that water emission is coincident with IRAS\,4A2 and a series of shocks in the NS direction while no water emission is seen to arise from IRAS\,4A1. In this scenario, hot corinos display significant emission from molecules that were inherited from the prestellar phase and survive frozen on the dust grains within their envelopes until they are heated and released in this phase for the first time. Deuterium fractionation and a very low ortho-to-para ratio of water in particular favors grain-surface formation at very low temperatures during the prestellar phase \citep{Ceccarelli:14a, Dionatos:20a}. Maintaining primordial material in their envelopes may indicate that such sources are younger compared to other YSOs seen in the field, which appear to have processed and exhausted the reservoir of primordial ices in their envelopes. This scenario is also consistent with the abundance increase in the inner envelope implied by models \citep{Mottram:13a} and can explain the exceptional high velocity water which is not observed in other sources \citep[e.g., Fig.~\ref{onlinefig:1},][]{Mottram:14a}.  

Assuming all volatile oxygen which is not locked up in CO, is bound in water, and that all water is in the gas phase, then the water abundance with respect to H$_2$ in thermodynamic equilibrium is estimated to be  $5 - 6 \times 10^{-4}$ \citep{vanDishoeck:14a}. Abundance estimations in this work are at best a factor of ten lower compared to the expected values but are in agreement with determinations based on a variety of other observations and methods \citep[e.g.,][]{Bjerkeli:12a, Tafalla:13a, Kristensen:17a}.  Water abundance estimations based on CO are a factor of $\sim$10 lower compared to the ones based on H$_2$ with the exception of the PDR around 03260+3111 where this relation is inverted mainly due to the very different responses of water and H$_2$ to UV heating (photodissociation vs self-shielding). In some cases water has been reported to closely follow the emission pattern of H$_2$ \citep{Tafalla:13a}, however, the spatial correlation between the two molecules is not clear in the maps presented here. In addition in the case UV-heating is present, then water abundances derived from comparisons with H$_2$ can be severely underestimated. In contrast, CO appears to be a better proxy for estimating water abundances as both molecules are affected in a similar way under the influence of UV radiation, while their resolved velocity profiles indicate that they trace the same volume of gas \citep{Kristensen:17a}.  

The main mechanism leading to lower water abundances than expected is not yet clear. As demonstrated in the PDR region, UV-heating can efficiently reduce water abundance and in fact, even modest radiation field strengths are sufficient to account for the observed abundance values \citep{Kristensen:17a}. The narrow component water map however confirms that UV heating may have an influence along HH 7-11 and  the outflow of IRAS4A  but not in other outflows. In these two particular cases the water abundance does not appear to vary significantly when compared to other outflows and therefore it is not clear what is the net effect of energetic radiation along the outflow cavities. On the other hand, UV heating may be ubiquitous in outflows but not traced in our maps due to the available sensitivity. Fast moving material in outflows can produce dissociative J-shocks, which seem to provide a convincing alternative that can explain the observed water abundance deficiency. The mass flux and momentum carried by water in outflows is comparable with the values measured for [\ion{O}{i}] that traces fast shocks, and this link provides additional support to the shock-dissociation scenario for water.

\section{Conclusions} \label{sec:6}

We presented Spitzer and Herschel spectral mapping observations of the NGC\,1333 star-forming region. Spitzer maps comprise of slit-scan observations using the long-slit IRS modules and record emission of molecular hydrogen. Herschel observations provide a velocity resolved map of water at 557\,GHz obtained with HIFI and a series of CO and [\ion{C}{i}] line maps observed with SPIRE. H$_2$ and CO emission was analyzed by means of excitation diagrams and mass flux was calculated for each molecule and outflow. The water mass flux was in addition calculated using a model derived description relating the water line intensity and the column density of the molecule. The velocity resolved lines for water were employed to calculate the momentum and energy of the outflows and compare it with previous estimations of other outflow tracers. Finally we calculated the abundance of water based on comparisons with CO and H$_2$ for all outflows and a few more positions of particular interest. The main results of our analysis can be summarized as follows:

\begin{itemize}

\item{Morphologically, molecular hydrogen traces emission related to sources of energetic radiation and, at lower levels, to outflows.}

\item{Water lines are resolved to a narrow (slow) and a wide (fast) component. When present, the narrow component contributes up to $\sim5\%$ of the total line luminosity. Its spatial distribution indicates that it is excited in regions where UV heating is active. The fast component appears to trace gas coasting in outflows after being accelerated in fast shocks as suggested from comparisons between the velocity centroid and integrated emission maps.}

\item{Integrated line emission of water peaks on IRAS\,4A showing at the same time an abundance increase of at least a factor of two compared to all other regions in NGC\,1333.  While outflow and UV-heating signatures are present in the line profile, these processes alone are not sufficient to explain the observed water enhancement. The fact that IRAS\,4A2 is a hot corino can provide reasonable grounds to interpret the difference in the water emission between IRAS\,4A and other protostars in NGC\,1333. The nature of IRAS\,4A2 is also confirmed with the serendipitous detection of acetone or methyl-formate in the HIFI maps.}

\item{SPIRE maps of carbon monoxide trace progressively deeper layers into NGC\,1333 with increasing upper level energy of the transitions. Emission becomes optically thin at $J_{up}\sim9$. Higher excitation CO transitions have similar morphological characteristics to the H$_2$ maps, being highly excited around the B-stars in the north but also tracing the complex web of outflows in the south. Intermediate J CO lines appear to peak at the locations where the narrow H$_2$O component becomes brighter, suggesting that a component of the intermediate J CO emission is sensitive to UV radiation. Radiative heating must be acting on small scales as it produces no observable differences in the excitation and kinematics between outflows with and without UV signatures.}

\item{A similar optical depth effect to CO is observed also for the two [\ion{C}{i}] emission lines detected with SPIRE, as the higher excitation transition appears to be excited around known protostellar sources. Temperature maps based on RADEX models show that [\ion{C}{i}] traces warm ($\sim$100K) material around protostellar envelopes.}

\item{Excitation analysis shows that H$_2$ traces warm ($\sim$500K) gas excited in outflows. A derived $A_v$ value of $\sim15$ mag along the outflows is consistent with previous estimations in the same region. Excitation of CO lines detected with SPIRE is best approximated with two temperatures of $\sim$ 50\,K and 100\,K while column densities range from $\sim 10^{19}$ cm$^{-2}$ for H$_2$ to $\sim 10^{16}$ cm$^{-2}$ for CO.}

\item{CO mass flux is measured to be an order of magnitude higher compared to H$_2$ and two orders of magnitude relative to H$_2$O. At the limit that the emission becomes optically thin, CO appears to trace the bulk of the outflow mass while H$_2$ and H$_2$O are probing the mass of the gas excited in shocks. The mass flux discrepancy between H$_2$ and H$_2$O can be explained by dissociation of water in shocks, which is also consistent with the low abundance measurements. This scenario is further supported by the observations that  the H$_2$O momentum and energy correlates well with the [\ion{O}{i}] measurements from \citetalias{Dionatos:17a}}


\item{Water abundance in outflows averages from  7\,$\times$\,10$^{-7}$ and 5\,$\times$\,10$^{-6}$, based on comparisons with CO and H$_2$, respectively.  CO- and H$_2$-based water abundances are a factor of 10 and 10$^{3}$ lower in the PDR around the B-star 03260+3111, primarily due to the different responses of molecules to energetic radiation.  Therefore UV heating can explain the lower water abundances, however the narrow H$_2$O component indicates that radiation may be active in only a couple of cases, and for those outflows significant abundance variations are observed. Dissociation of water in shocks on the other hand is consistent with the results of the morphological, kinematical and excitation analysis and provides the only alternative pathway to interpret the observed water abundance deficiency. }

\item{CO at the limit that it becomes optically thin is a safer proxy to estimate water abundances than H$_2$, as it has a closer response to UV fields and traces the same velocity components in outflows as water.}





\end{itemize}

The observations presented here but also in \citepalias{Dionatos:17a} demonstrate the power of large-scale maps in interpreting emission from complex systems such as low-mass protostars in a consistent way. They also underline the importance of such maps as opposed to pointed observations in performing consistent comparisons between sources and identifying the ones that appear to have particular properties. While the role of the various molecules in tracing diverse processes starts taking shape, observations of mid-infrared and far infrared lines at higher spectral and angular resolution in the future will be important to understand the detailed mechanisms governing the excitation and cooling of protostellar jets. To this end, observations with SOFIA and soon JWST along with the future missions such as SPICA will play an important role in deciphering the role of protostellar ejecta as a feedback mechanism in star-formation.

\begin{acknowledgements}
 This research was supported by the Austrian Research Promotion Agency (FFG)  under the framework of the Austrian Space Applications Program (ASAP) project PROTEUS (FFG-866005). The research of LEK is supported by a research grant (19127) from VILLUM FONDEN. Research at the Centre for Star and Planet Formation is funded by the Danish National Research Foundation.
\end{acknowledgements}

%
%


\begin{tiny}

\bibliographystyle{aa}
\bibliography{NGC1333_spire_hifi}

\begin{thebibliography}{42}
\expandafter\ifx\csname natexlab\endcsname\relax\def\natexlab#1{#1}\fi

\bibitem[{{Bergin} {et~al.}(2003){Bergin}, {Kaufman}, {Melnick}, {Snell}, \&
  {Howe}}]{Bergin:03a}
{Bergin}, E.~A., {Kaufman}, M.~J., {Melnick}, G.~J., {Snell}, R.~L., \& {Howe},
  J.~E. 2003, \apj, 582, 830

\bibitem[{{Bjerkeli} {et~al.}(2012){Bjerkeli}, {Liseau}, {Larsson}, {Rydbeck},
  {Nisini}, {Tafalla}, {Antoniucci}, {Benedettini}, {Bergman}, {Cabrit},
  {Giannini}, {Melnick}, {Neufeld}, {Santangelo}, \& {van
  Dishoeck}}]{Bjerkeli:12a}
{Bjerkeli}, P., {Liseau}, R., {Larsson}, B., {et~al.} 2012, \aap, 546, A29

\bibitem[{{Bottinelli} {et~al.}(2004){Bottinelli}, {Ceccarelli}, {Lefloch},
  {Williams}, {Castets}, {Caux}, {Cazaux}, {Maret}, {Parise}, \&
  {Tielens}}]{Bottinelli:04a}
{Bottinelli}, S., {Ceccarelli}, C., {Lefloch}, B., {et~al.} 2004, \apj, 615,
  354

\bibitem[{{Ceccarelli} {et~al.}(2014){Ceccarelli}, {Caselli},
  {Bockel{\'e}e-Morvan}, {Mousis}, {Pizzarello}, {Robert}, \&
  {Semenov}}]{Ceccarelli:14a}
{Ceccarelli}, C., {Caselli}, P., {Bockel{\'e}e-Morvan}, D., {et~al.} 2014, in
  Protostars and Planets VI, ed. H.~{Beuther}, R.~S. {Klessen}, C.~P.
  {Dullemond}, \& T.~{Henning}, 859

\bibitem[{{Chapman} {et~al.}(2009){Chapman}, {Mundy}, {Lai}, \&
  {Evans}}]{Chapman:09a}
{Chapman}, N.~L., {Mundy}, L.~G., {Lai}, S.-P., \& {Evans}, II, N.~J. 2009,
  \apj, 690, 496

\bibitem[{{de Graauw} {et~al.}(2010){de Graauw}, {Helmich}, {Phillips},
  {Stutzki}, {Caux}, {Whyborn}, {Dieleman}, {Roelfsema}, {Aarts}, {Assendorp},
  {Bachiller}, {Baechtold}, {Barcia}, {Beintema}, {Belitsky}, {Benz}, {Bieber},
  {Boogert}, {Borys}, {Bumble}, {Ca{\"i}s}, {Caris}, {Cerulli-Irelli},
  {Chattopadhyay}, {Cherednichenko}, {Ciechanowicz}, {Coeur-Joly}, {Comito},
  {Cros}, {de Jonge}, {de Lange}, {Delforges}, {Delorme}, {den Boggende},
  {Desbat}, {Diez-Gonz{\'a}lez}, {di Giorgio}, {Dubbeldam}, {Edwards},
  {Eggens}, {Erickson}, {Evers}, {Fich}, {Finn}, {Franke}, {Gaier}, {Gal},
  {Gao}, {Gallego}, {Gauffre}, {Gill}, {Glenz}, {Golstein}, {Goulooze},
  {Gunsing}, {G{\"u}sten}, {Hartogh}, {Hatch}, {Higgins}, {Honingh}, {Huisman},
  {Jackson}, {Jacobs}, {Jacobs}, {Jarchow}, {Javadi}, {Jellema}, {Justen},
  {Karpov}, {Kasemann}, {Kawamura}, {Keizer}, {Kester}, {Klapwijk}, {Klein},
  {Kollberg}, {Kooi}, {Kooiman}, {Kopf}, {Krause}, {Krieg}, {Kramer},
  {Kruizenga}, {Kuhn}, {Laauwen}, {Lai}, {Larsson}, {Leduc}, {Leinz}, {Lin},
  {Liseau}, {Liu}, {Loose}, {L{\'o}pez-Fernandez}, {Lord}, {Luinge}, {Marston},
  {Mart{\'{\i}}n-Pintado}, {Maestrini}, {Maiwald}, {McCoey}, {Mehdi}, {Megej},
  {Melchior}, {Meinsma}, {Merkel}, {Michalska}, {Monstein}, {Moratschke},
  {Morris}, {Muller}, {Murphy}, {Naber}, {Natale}, {Nowosielski}, {Nuzzolo},
  {Olberg}, {Olbrich}, {Orfei}, {Orleanski}, {Ossenkopf}, {Peacock}, {Pearson},
  {Peron}, {Phillip-May}, {Piazzo}, {Planesas}, {Rataj}, {Ravera}, {Risacher},
  {Salez}, {Samoska}, {Saraceno}, {Schieder}, {Schlecht}, {Schl{\"o}der},
  {Schm{\"u}lling}, {Schultz}, {Schuster}, {Siebertz}, {Smit}, {Szczerba},
  {Shipman}, {Steinmetz}, {Stern}, {Stokroos}, {Teipen}, {Teyssier}, {Tils},
  {Trappe}, {van Baaren}, {van Leeuwen}, {van de Stadt}, {Visser}, {Wildeman},
  {Wafelbakker}, {Ward}, {Wesselius}, {Wild}, {Wulff}, {Wunsch}, {Tielens},
  {Zaal}, {Zirath}, {Zmuidzinas}, \& {Zwart}}]{deGraauw:10a}
{de Graauw}, T., {Helmich}, F.~P., {Phillips}, T.~G., {et~al.} 2010, \aap, 518,
  L6

\bibitem[{{Dionatos}(2020)}]{Dionatos:20a}
{Dionatos}, O. 2020, in IAU Symposium, Vol. 345, IAU Symposium, ed. B.~G.
  {Elmegreen}, L.~V. {T{\'o}th}, \& M.~{G{\"u}del}, 252--254

\bibitem[{{Dionatos} \& {G{\"u}del}(2017)}]{Dionatos:17a}
{Dionatos}, O. \& {G{\"u}del}, M. 2017, \aap, 597, A64

\bibitem[{{Dionatos} {et~al.}(2013){Dionatos}, {J{\o}rgensen}, {Green},
  {Herczeg}, {Evans}, {Kristensen}, {Lindberg}, \& {van
  Dishoeck}}]{Dionatos:13a}
{Dionatos}, O., {J{\o}rgensen}, J.~K., {Green}, J.~D., {et~al.} 2013, \aap,
  558, A88

\bibitem[{{Dionatos} {et~al.}(2014){Dionatos}, {J{\o}rgensen}, {Teixeira},
  {G{\"u}del}, \& {Bergin}}]{Dionatos:14a}
{Dionatos}, O., {J{\o}rgensen}, J.~K., {Teixeira}, P.~S., {G{\"u}del}, M., \&
  {Bergin}, E. 2014, \aap, 563, A28

\bibitem[{{Dionatos} {et~al.}(2010{\natexlab{a}}){Dionatos}, {Nisini},
  {Cabrit}, {Kristensen}, \& {Pineau Des For{\^e}ts}}]{Dionatos:10a}
{Dionatos}, O., {Nisini}, B., {Cabrit}, S., {Kristensen}, L., \& {Pineau Des
  For{\^e}ts}, G. 2010{\natexlab{a}}, \aap, 521, A7

\bibitem[{{Dionatos} {et~al.}(2010{\natexlab{b}}){Dionatos}, {Nisini},
  {Codella}, \& {Giannini}}]{Dionatos:10b}
{Dionatos}, O., {Nisini}, B., {Codella}, C., \& {Giannini}, T.
  2010{\natexlab{b}}, \aap, 523, A29

\bibitem[{{Dionatos} {et~al.}(2009){Dionatos}, {Nisini}, {Garcia Lopez},
  {Giannini}, {Davis}, {Smith}, {Ray}, \& {DeLuca}}]{Dionatos:09a}
{Dionatos}, O., {Nisini}, B., {Garcia Lopez}, R., {et~al.} 2009, \apj, 692, 1

\bibitem[{{Dionatos} {et~al.}(2018){Dionatos}, {Ray}, \&
  {G{\"u}del}}]{Dionatos:18a}
{Dionatos}, O., {Ray}, T., \& {G{\"u}del}, M. 2018, \aap, 616, A84

\bibitem[{{Green} {et~al.}(2013){Green}, {Evans}, {J{\o}rgensen}, {Herczeg},
  {Kristensen}, {Lee}, {Dionatos}, {Yildiz}, {Salyk}, {Meeus}, {Bouwman},
  {Visser}, {Bergin}, {van Dishoeck}, {Rascati}, {Karska}, {van Kempen},
  {Dunham}, {Lindberg}, {Fedele}, \& {DIGIT Team}}]{Green:13a}
{Green}, J.~D., {Evans}, II, N.~J., {J{\o}rgensen}, J.~K., {et~al.} 2013, \apj,
  770, 123

\bibitem[{{Griffin} {et~al.}(2010){Griffin}, {Abergel}, {Abreu}, {Ade},
  {Andr{\'e}}, {Augueres}, {Babbedge}, {Bae}, {Baillie}, {Baluteau}, {Barlow},
  {Bendo}, {Benielli}, {Bock}, {Bonhomme}, {Brisbin}, {Brockley-Blatt},
  {Caldwell}, {Cara}, {Castro-Rodriguez}, {Cerulli}, {Chanial}, {Chen},
  {Clark}, {Clements}, {Clerc}, {Coker}, {Communal}, {Conversi}, {Cox},
  {Crumb}, {Cunningham}, {Daly}, {Davis}, {de Antoni}, {Delderfield}, {Devin},
  {di Giorgio}, {Didschuns}, {Dohlen}, {Donati}, {Dowell}, {Dowell}, {Duband},
  {Dumaye}, {Emery}, {Ferlet}, {Ferrand}, {Fontignie}, {Fox}, {Franceschini},
  {Frerking}, {Fulton}, {Garcia}, {Gastaud}, {Gear}, {Glenn}, {Goizel},
  {Griffin}, {Grundy}, {Guest}, {Guillemet}, {Hargrave}, {Harwit}, {Hastings},
  {Hatziminaoglou}, {Herman}, {Hinde}, {Hristov}, {Huang}, {Imhof}, {Isaak},
  {Israelsson}, {Ivison}, {Jennings}, {Kiernan}, {King}, {Lange}, {Latter},
  {Laurent}, {Laurent}, {Leeks}, {Lellouch}, {Levenson}, {Li}, {Li},
  {Lilienthal}, {Lim}, {Liu}, {Lu}, {Madden}, {Mainetti}, {Marliani}, {McKay},
  {Mercier}, {Molinari}, {Morris}, {Moseley}, {Mulder}, {Mur}, {Naylor},
  {Nguyen}, {O'Halloran}, {Oliver}, {Olofsson}, {Olofsson}, {Orfei}, {Page},
  {Pain}, {Panuzzo}, {Papageorgiou}, {Parks}, {Parr-Burman}, {Pearce},
  {Pearson}, {P{\'e}rez-Fournon}, {Pinsard}, {Pisano}, {Podosek}, {Pohlen},
  {Polehampton}, {Pouliquen}, {Rigopoulou}, {Rizzo}, {Roseboom}, {Roussel},
  {Rowan-Robinson}, {Rownd}, {Saraceno}, {Sauvage}, {Savage}, {Savini},
  {Sawyer}, {Scharmberg}, {Schmitt}, {Schneider}, {Schulz}, {Schwartz},
  {Shafer}, {Shupe}, {Sibthorpe}, {Sidher}, {Smith}, {Smith}, {Smith},
  {Spencer}, {Stobie}, {Sudiwala}, {Sukhatme}, {Surace}, {Stevens}, {Swinyard},
  {Trichas}, {Tourette}, {Triou}, {Tseng}, {Tucker}, {Turner}, {Vaccari},
  {Valtchanov}, {Vigroux}, {Virique}, {Voellmer}, {Walker}, {Ward}, {Waskett},
  {Weilert}, {Wesson}, {White}, {Whitehouse}, {Wilson}, {Winter}, {Woodcraft},
  {Wright}, {Xu}, {Zavagno}, {Zemcov}, {Zhang}, \& {Zonca}}]{Griffin:10a}
{Griffin}, M.~J., {Abergel}, A., {Abreu}, A., {et~al.} 2010, \aap, 518, L3

\bibitem[{{Gutermuth} {et~al.}(2008){Gutermuth}, {Myers}, {Megeath}, {Allen},
  {Pipher}, {Muzerolle}, {Porras}, {Winston}, \& {Fazio}}]{Gutermuth:08a}
{Gutermuth}, R.~A., {Myers}, P.~C., {Megeath}, S.~T., {et~al.} 2008, \apj, 674,
  336

\bibitem[{{Herczeg} {et~al.}(2012){Herczeg}, {Karska}, {Bruderer},
  {Kristensen}, {van Dishoeck}, {J{\o}rgensen}, {Visser}, {Wampfler}, {Bergin},
  \& {Y{\i}ld{\i}z}}]{Herczeg:12a}
{Herczeg}, G.~J., {Karska}, A., {Bruderer}, S., {et~al.} 2012, \aap, 540, A84

\bibitem[{{Hollenbach} \& {Tielens}(1997)}]{Hollenbach:97a}
{Hollenbach}, D.~J. \& {Tielens}, A.~G.~G.~M. 1997, \araa, 35, 179

\bibitem[{{Karska} {et~al.}(2018){Karska}, {Kaufman}, {Kristensen}, {van
  Dishoeck}, {Herczeg}, {Mottram}, {Tychoniec}, {Lindberg}, {Evans}, {Green},
  {Yang}, {Gusdorf}, {Itrich}, \& {Si{\'o}dmiak}}]{Karska:18a}
{Karska}, A., {Kaufman}, M.~J., {Kristensen}, L.~E., {et~al.} 2018, \apjs, 235,
  30

\bibitem[{{Kristensen} {et~al.}(2012){Kristensen}, {van Dishoeck}, {Bergin},
  {Visser}, {Y{\i}ld{\i}z}, {San Jose-Garcia}, {J{\o}rgensen}, {Herczeg},
  {Johnstone}, {Wampfler}, {Benz}, {Bruderer}, {Cabrit}, {Caselli}, {Doty},
  {Harsono}, {Herpin}, {Hogerheijde}, {Karska}, {van Kempen}, {Liseau},
  {Nisini}, {Tafalla}, {van der Tak}, \& {Wyrowski}}]{Kristensen:12a}
{Kristensen}, L.~E., {van Dishoeck}, E.~F., {Bergin}, E.~A., {et~al.} 2012,
  \aap, 542, A8

\bibitem[{{Kristensen} {et~al.}(2017){Kristensen}, {van Dishoeck}, {Mottram},
  {Karska}, {Y{\i}ld{\i}z}, {Bergin}, {Bjerkeli}, {Cabrit}, {Doty}, {Evans},
  {Gusdorf}, {Harsono}, {Herczeg}, {Johnstone}, {J{\o}rgensen}, {van Kempen},
  {Lee}, {Maret}, {Tafalla}, {Visser}, \& {Wampfler}}]{Kristensen:17a}
{Kristensen}, L.~E., {van Dishoeck}, E.~F., {Mottram}, J.~C., {et~al.} 2017,
  \aap, 605, A93

\bibitem[{{Maret} {et~al.}(2009){Maret}, {Bergin}, {Neufeld}, {Green},
  {Watson}, {Harwit}, {Kristensen}, {Melnick}, {Sonnentrucker}, {Tolls},
  {Werner}, {Willacy}, \& {Yuan}}]{Maret:09a}
{Maret}, S., {Bergin}, E.~A., {Neufeld}, D.~A., {et~al.} 2009, \apj, 698, 1244

\bibitem[{{Mottram} {et~al.}(2014){Mottram}, {Kristensen}, {van Dishoeck},
  {Bruderer}, {San Jos{\'e}-Garc{\'\i}a}, {Karska}, {Visser}, {Santangelo},
  {Benz}, \& {Bergin}}]{Mottram:14a}
{Mottram}, J.~C., {Kristensen}, L.~E., {van Dishoeck}, E.~F., {et~al.} 2014,
  \aap, 572, A21

\bibitem[{{Mottram} {et~al.}(2013){Mottram}, {van Dishoeck}, {Schmalzl},
  {Kristensen}, {Visser}, {Hogerheijde}, \& {Bruderer}}]{Mottram:13a}
{Mottram}, J.~C., {van Dishoeck}, E.~F., {Schmalzl}, M., {et~al.} 2013, \aap,
  558, A126

\bibitem[{{Persson} {et~al.}(2012){Persson}, {J{\o}rgensen}, \& {van
  Dishoeck}}]{Persson:12a}
{Persson}, M.~V., {J{\o}rgensen}, J.~K., \& {van Dishoeck}, E.~F. 2012, \aap,
  541, A39

\bibitem[{{Pilbratt} {et~al.}(2010){Pilbratt}, {Riedinger}, {Passvogel},
  {Crone}, {Doyle}, {Gageur}, {Heras}, {Jewell}, {Metcalfe}, {Ott}, \&
  {Schmidt}}]{Pilbratt:10a}
{Pilbratt}, G.~L., {Riedinger}, J.~R., {Passvogel}, T., {et~al.} 2010, \aap,
  518, L1

\bibitem[{{Plunkett} {et~al.}(2013){Plunkett}, {Arce}, {Corder}, {Mardones},
  {Sargent}, \& {Schnee}}]{Plunkett:13a}
{Plunkett}, A.~L., {Arce}, H.~G., {Corder}, S.~A., {et~al.} 2013, \apj, 774, 22

\bibitem[{{Raga} {et~al.}(2013){Raga}, {Noriega-Crespo}, {Carey}, \&
  {Arce}}]{Raga:13a}
{Raga}, A.~C., {Noriega-Crespo}, A., {Carey}, S.~J., \& {Arce}, H.~G. 2013,
  \aj, 145, 28

\bibitem[{{Sahu} {et~al.}(2019){Sahu}, {Liu}, {Su}, {Li}, {Lee}, {Hirano}, \&
  {Takakuwa}}]{Sahu:19a}
{Sahu}, D., {Liu}, S.-Y., {Su}, Y.-N., {et~al.} 2019, \apj, 872, 196

\bibitem[{{Schroder} {et~al.}(1991){Schroder}, {Staemmler}, {Smith}, {Flower},
  \& {Jaquet}}]{Schroder:91a}
{Schroder}, K., {Staemmler}, V., {Smith}, M.~D., {Flower}, D.~R., \& {Jaquet},
  R. 1991, Journal of Physics B Atomic Molecular Physics, 24, 2487

\bibitem[{{Smith} {et~al.}(2007){Smith}, {Armus}, {Dale}, {Roussel}, {Sheth},
  {Buckalew}, {Jarrett}, {Helou}, \& {Kennicutt}}]{Smith:07a}
{Smith}, J.~D.~T., {Armus}, L., {Dale}, D.~A., {et~al.} 2007, \pasp, 119, 1133

\bibitem[{{Spaans} {et~al.}(1995){Spaans}, {Hogerheijde}, {Mundy}, \& {van
  Dishoeck}}]{Spaans:95a}
{Spaans}, M., {Hogerheijde}, M.~R., {Mundy}, L.~G., \& {van Dishoeck}, E.~F.
  1995, \apjl, 455, L167

\bibitem[{{Suutarinen} {et~al.}(2014){Suutarinen}, {Kristensen}, {Mottram},
  {Fraser}, \& {van Dishoeck}}]{Suutarinen:14a}
{Suutarinen}, A.~N., {Kristensen}, L.~E., {Mottram}, J.~C., {Fraser}, H.~J., \&
  {van Dishoeck}, E.~F. 2014, \mnras, 440, 1844

\bibitem[{{Tafalla} {et~al.}(2013){Tafalla}, {Liseau}, {Nisini}, {Bachiller},
  {Santiago-Garc{\'{\i}}a}, {van Dishoeck}, {Kristensen}, {Herczeg}, \&
  {Y{\i}ld{\i}z}}]{Tafalla:13a}
{Tafalla}, M., {Liseau}, R., {Nisini}, B., {et~al.} 2013, \aap, 551, A116

\bibitem[{{Tappe} {et~al.}(2008){Tappe}, {Lada}, {Black}, \&
  {Muench}}]{Tappe:09a}
{Tappe}, A., {Lada}, C.~J., {Black}, J.~H., \& {Muench}, A.~A. 2008, \apjl,
  680, L117

\bibitem[{{van der Tak} {et~al.}(2007){van der Tak}, {Black}, {Sch{\"o}ier},
  {Jansen}, \& {van Dishoeck}}]{vanderTak:07a}
{van der Tak}, F.~F.~S., {Black}, J.~H., {Sch{\"o}ier}, F.~L., {Jansen}, D.~J.,
  \& {van Dishoeck}, E.~F. 2007, \aap, 468, 627

\bibitem[{{van Dishoeck} {et~al.}(2014){van Dishoeck}, {Bergin}, {Lis}, \&
  {Lunine}}]{vanDishoeck:14a}
{van Dishoeck}, E.~F., {Bergin}, E.~A., {Lis}, D.~C., \& {Lunine}, J.~I. 2014,
  Protostars and Planets VI, 835

\bibitem[{{Walawender} {et~al.}(2008){Walawender}, {Bally}, {Francesco},
  {J{\o}rgensen}, \& {Getman}}]{Walawender:08a}
{Walawender}, J., {Bally}, J., {Francesco}, J.~D., {J{\o}rgensen}, J., \&
  {Getman}, K.~. 2008, {NGC 1333: A Nearby Burst of Star Formation}, ed.
  B.~{Reipurth}, 346

\bibitem[{{Yang} {et~al.}(2018){Yang}, {Green}, {Evans}, {Lee}, {J{\o}rgensen},
  {Kristensen}, {Mottram}, {Herczeg}, {Karska}, {Dionatos}, {Bergin},
  {Bouwman}, {van Dishoeck}, {van Kempen}, {Larson}, \&
  {Y{\i}ld{\i}z}}]{Yang:18a}
{Yang}, Y.-L., {Green}, J.~D., {Evans}, II, N.~J., {et~al.} 2018, \apj, 860,
  174

\bibitem[{{Y{\i}ld{\i}z} {et~al.}(2012){Y{\i}ld{\i}z}, {Kristensen}, {van
  Dishoeck}, {Belloche}, {van Kempen}, {Hogerheijde}, {G{\"u}sten}, \& {van der
  Marel}}]{Yildiz:12a}
{Y{\i}ld{\i}z}, U.~A., {Kristensen}, L.~E., {van Dishoeck}, E.~F., {et~al.}
  2012, \aap, 542, A86

\bibitem[{{Y{\i}ld{\i}z} {et~al.}(2013){Y{\i}ld{\i}z}, {Kristensen}, {van
  Dishoeck}, {San Jos{\'e}-Garc{\'\i}a}, {Karska}, {Harsono}, {Tafalla},
  {Fuente}, {Visser}, {J{\o}rgensen}, \& {Hogerheijde}}]{Yildiz:13a}
{Y{\i}ld{\i}z}, U.~A., {Kristensen}, L.~E., {van Dishoeck}, E.~F., {et~al.}
  2013, \aap, 556, A89

\end{thebibliography}

\end{tiny}


%

\begin{appendix} 
\section{Additional maps and figures}
\label{app:A}

\begin{figure}[!h]
\centering
\includegraphics[width=17.9cm]{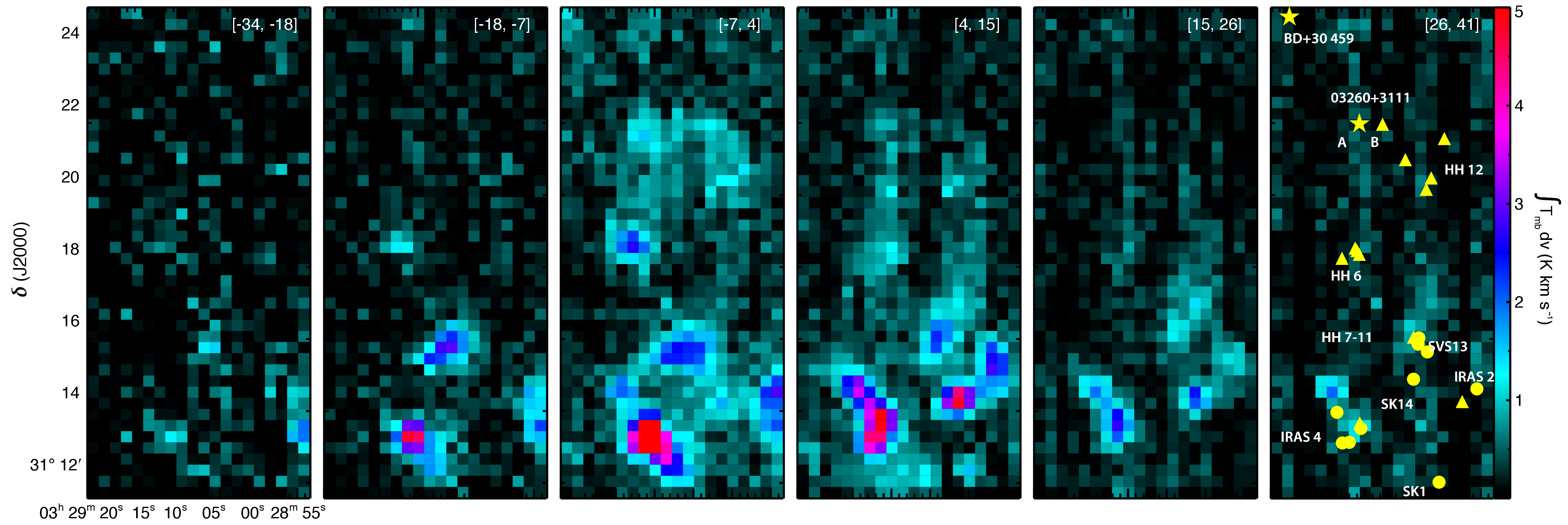}
\caption {Channel maps of the water line displaying the integrated emission at different velocity intervals which are indicated on the top right corner of each panel.} 
\label{onlinefig:1}
\end{figure} 

\begin{figure}[!h]
\centering
\includegraphics[width=8.8cm]{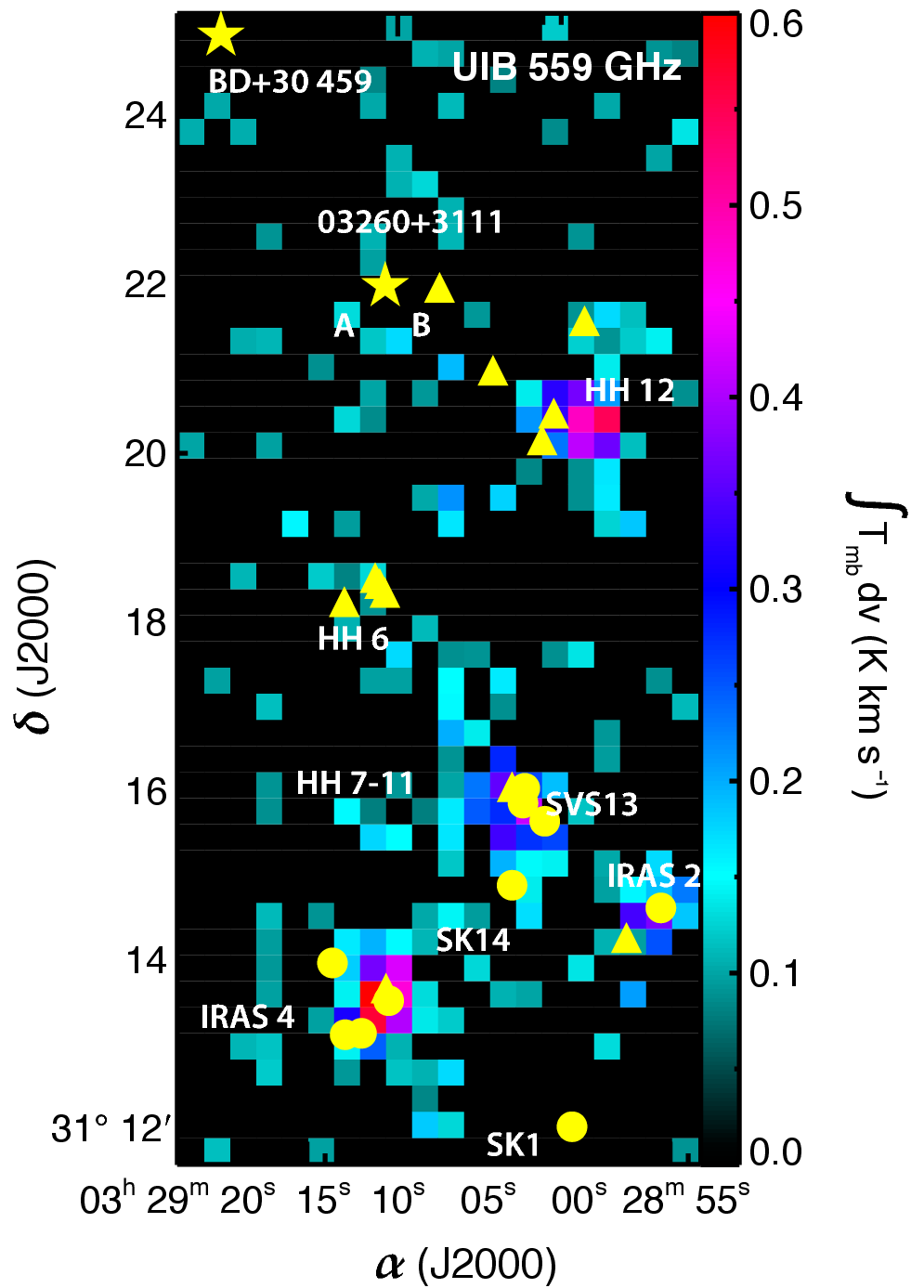}
\caption {Map of a narrow line, possibly from either acetone ((CH$_3$)$_2$CO) or methyl formate (CH$_3$OCHO). Being excited in the vicinity of embedded sources, it may trace hot-corino activity. } 
\label{onlinefig:2}
\end{figure}


\begin{figure*}[!t]
\centering
\includegraphics[width=17.9cm]{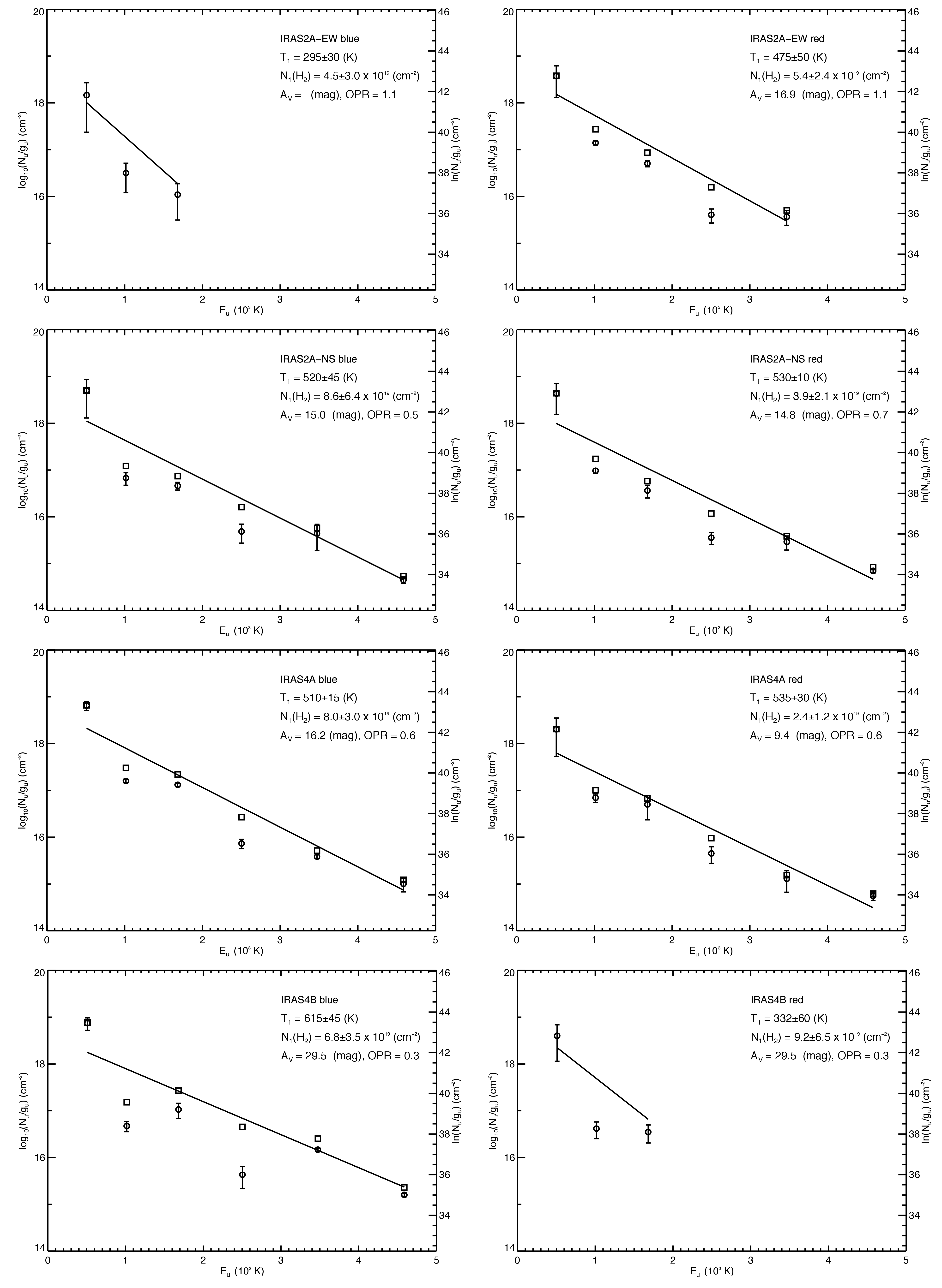}
\caption {Excitation diagrams for the H$_2$ fluxes extracted from Spitzer within the elliptical regions. } 
\label{onlinefig:3}
\end{figure*} 

\begin{figure*}
\centering
\includegraphics[width=17.9cm]{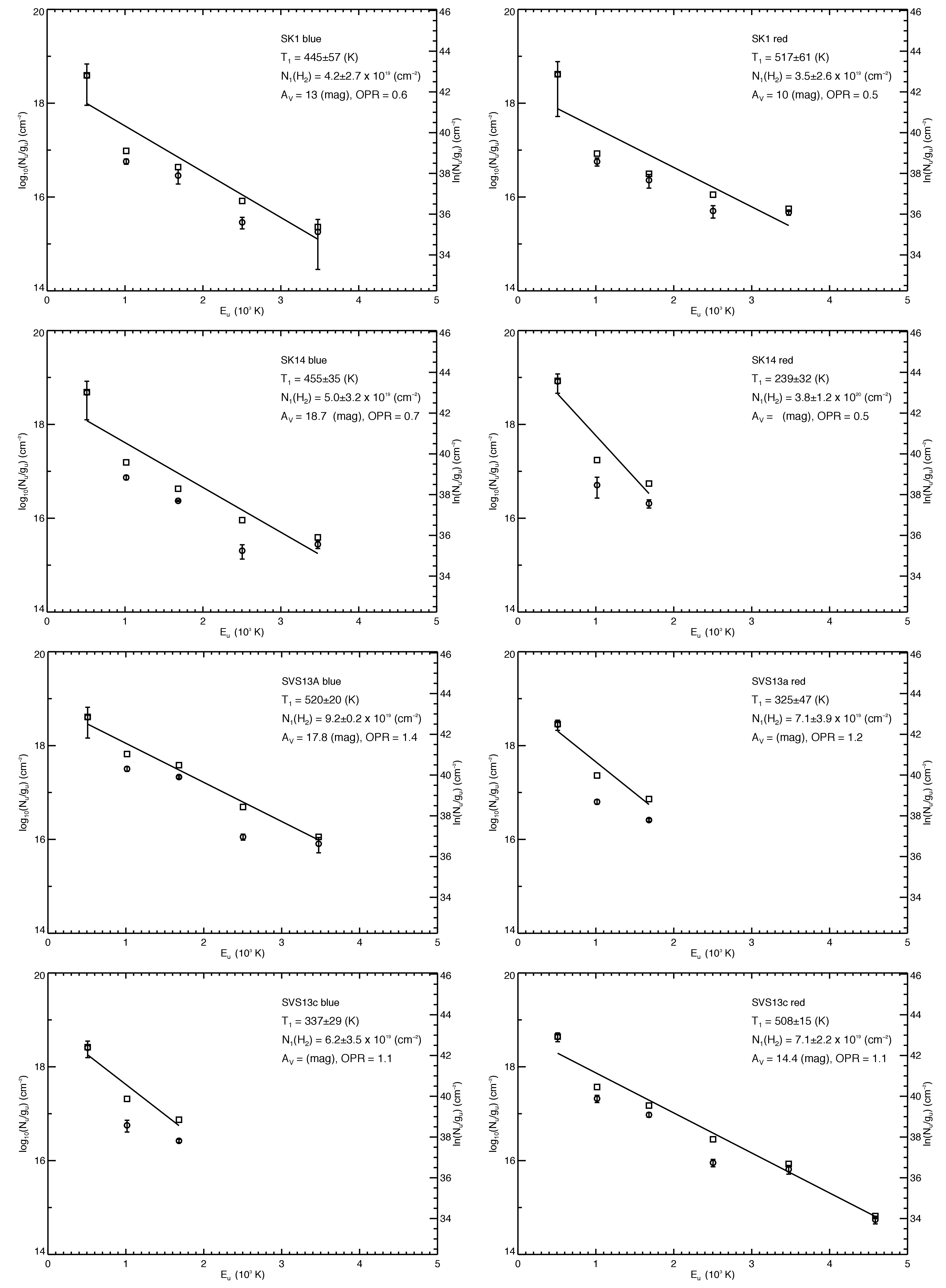}
\caption {Excitation diagrams for the H$_2$ fluxes extracted from Spitzer within the elliptical regions. } 
\label{onlinefig:4}
\end{figure*} 


\begin{figure*}
\centering
\includegraphics[width=17.9cm]{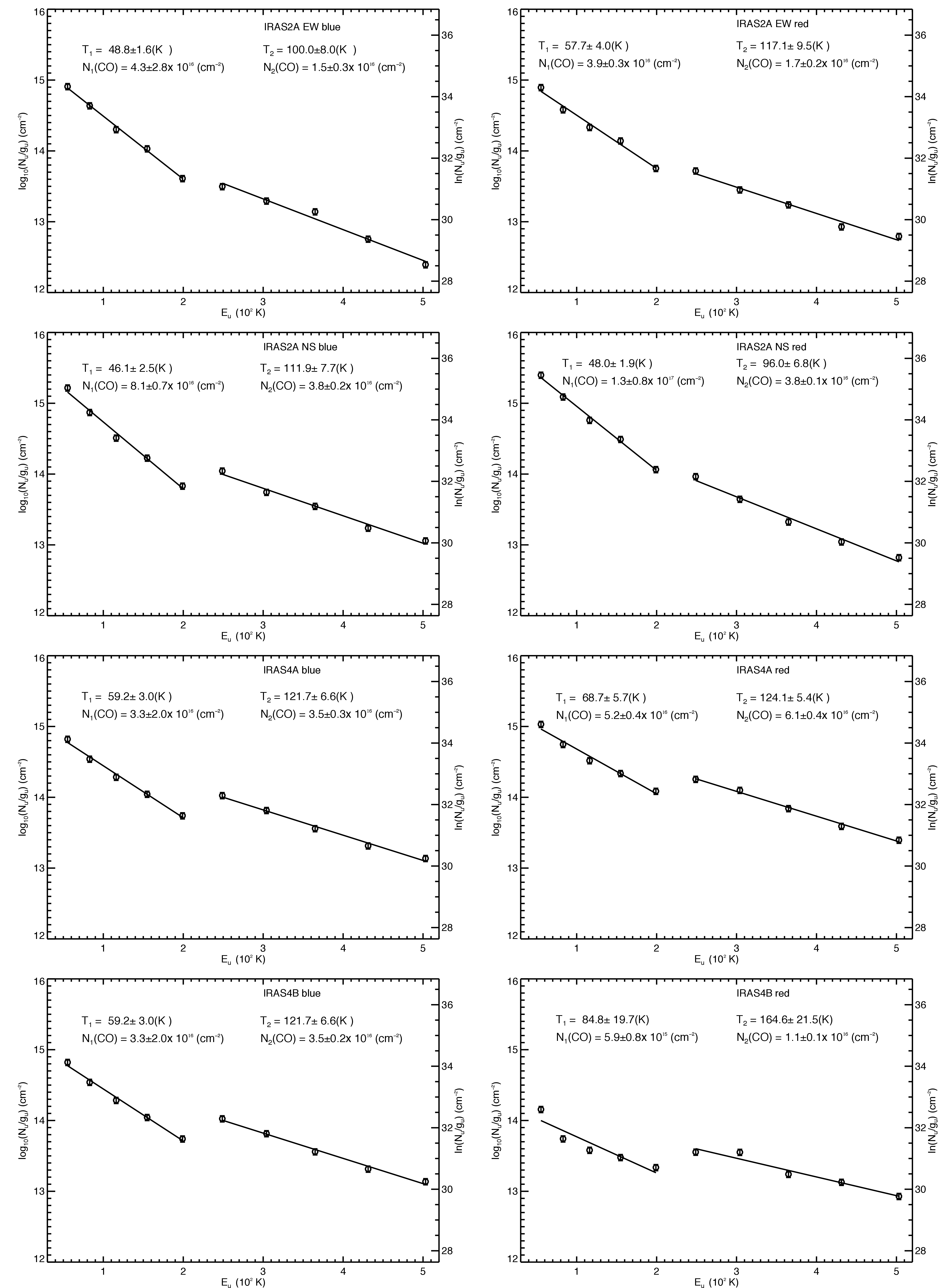}
\caption {Excitation diagrams for CO fluxes extracted from the SPIRE maps } 
\label{onlinefig:5}
\end{figure*}

 \begin{figure*}
\centering
\includegraphics[width=17.9cm]{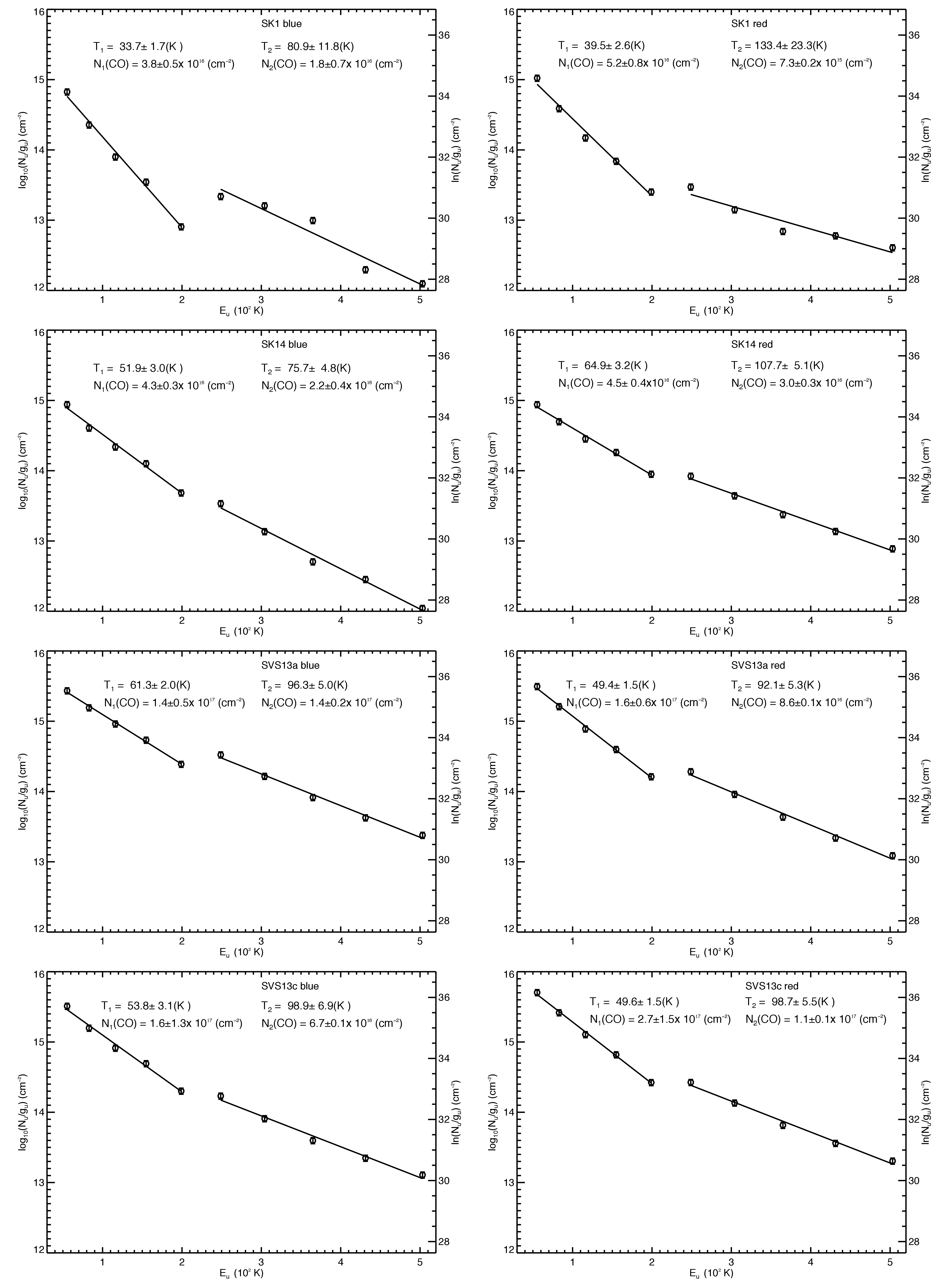}
\caption {Excitation diagrams for CO fluxes extracted from the SPIRE maps } 
\label{onlinefig:6}
\end{figure*}


\end{appendix}
\end{document}